\documentclass[reprint,superscriptaddress,amsmath,amssymb,longbibliography,prl]{revtex4-2}

\usepackage{amsmath}
\usepackage{mathtools}
\usepackage{amssymb}
\usepackage{graphicx}
\usepackage{bm}
\usepackage[colorlinks, linkcolor=blue, citecolor=blue, urlcolor=blue, breaklinks=true]{hyperref}
\usepackage{color,dsfont,multirow,booktabs}
\usepackage{braket}
\usepackage{tabularx}
\usepackage{lipsum,booktabs}
\usepackage{subfiles}
\usepackage{caption}
\usepackage[dvipsnames]{xcolor}
\usepackage{ragged2e}
\usepackage{orcidlink}
\usepackage[T1]{fontenc}
\usepackage[normalem]{ulem}

\newcommand{\commentold}[1]{}
\DeclareMathSymbol{:}{\mathpunct}{operators}{"3A}

\newcommand{\figpanel}[2]{\hyperref[#1]{\ref*{#1}(#2)}}

\begin{document}
\date{\today}

\title{Phase-Locked Quantum Battery Charging via Dissipative Mode Selection}

\author{Borhan Ahmadi\orcidlink{0000-0002-2787-9321}}
\address{International Centre for Theory of Quantum Technologies, University of Gdańsk, ul. prof. Marii Janion 4, 80-309 Gdańsk, Poland}

\author{Lei Du \orcidlink{0000-0003-0641-440X}}
\email{lei.du@chalmers.se}
\affiliation{Department of Microtechnology and Nanoscience (MC2), Chalmers University of Technology, 412 96 Gothenburg, Sweden}

\author{André H. A. Malavazi\orcidlink{0000-0002-0280-0621}}
\address{International Centre for Theory of Quantum Technologies, University of Gdańsk, ul. prof. Marii Janion 4, 80-309 Gdańsk, Poland}

\author{Shabir Barzanjeh}
\address{Department of Physics and Astronomy, University of Calgary, Calgary, AB T2N 1N4 Canada}

\author{Adam Miranowicz}
\address{Institute of Spintronics and Quantum Information,
Faculty of Physics and Astronomy, Adam Mickiewicz University,
61-614 Pozna\'n, Poland}

\author{Paweł Horodecki}
\address{International Centre for Theory of Quantum Technologies, University of Gdańsk, ul. prof. Marii Janion 4, 80-309 Gdańsk, Poland}

\begin{abstract}
We investigate how dynamical instability and reservoir engineering
can be harnessed to enhance the quality of quantum-battery
charging in a bosonic two-mode system. We show that reservoir-engineered dissipative coupling selectively amplifies a
single unstable quasi-dark mode, enabling phase-locked exponential
charging at a reduced gain threshold, whereas coherent coupling
can amplify two frequency-split unstable modes, resulting in
persistent beating. This dissipative mode selection yields a
near-unit ergotropy fraction and a substantially higher work-like signal-to-noise ratio (SNR), defined as ergotropy divided by the standard deviation of battery energy. Although optimizing the driving frequency enables coherent coupling to achieve faster charging and higher ergotropy power, dissipative coupling retains its advantage in the work-like SNR. An exact finite-time source–noise relation reveals that this advantage arises from a more constructive accumulation of coherent driving relative to
incoherent fluctuations. Our results establish dissipative mode selection as a promising route to high-quality quantum-battery charging.
\end{abstract}

\maketitle


\textit{Introduction.}---Gain plays a crucial role in sustaining and amplifying excitations in dissipative systems~\cite{ElGanainy2018}. It makes weak microwave signals measurable in quantum circuits, sustains lasers and masers, and compensates for propagation losses in optical systems~\cite{Macklin2015,Astafiev2007,Desurvire1987}. However, amplification also introduces fluctuations, so a stronger output does not necessarily mean a more useful one~\cite{Caves1982,Clerk2010}. This distinction becomes particularly important for quantum batteries~\cite{Campaioli2024Colloquium,ferraro2026opportunities}, where stored energy and extractable work need not coincide. The maximum extractable work is quantified by ergotropy, while the remaining stored energy is passive~\cite{Allahverdyan2004,Lobejko2022workfluctuations,2jtp-jpkn,ahmadi2026Amplifier,anuradha2026unequal}. For gain-assisted charging, the challenge is therefore to increase extractable work while controlling the accompanying energy fluctuations~\cite{GelbwaserKlimovsky2013,FriisHuber2018,PhysRevLett.125.040601,Crescente_2020,e24060820,6kwv-z6fx,PhysRevE.109.014131,Mohan2026,AhmadiChiral2026,ParametricQB2026}.

Engineered dissipation offers a promising way to address this challenge. A properly designed common reservoir can reshape the collective modes of a system and control their response to driving and noise~\cite{Diehl2008,ReservoirEngineering2014,Barzanjeh2026, ReservoirEngineering2015}. In coupled oscillators, different decay rates of collective modes can lead to a long-lived mode and promote synchronized dynamics~\cite{Giorgi2012, Manzano2013}. Related approaches have been explored to facilitate energy transfer and quantum-battery charging~\cite{Barra2019,Quach2020Dark,PhysRevLett.132.210402,Ahmadi2025,Pokhrel2025,Shastri2025,SubwavelengthQB2025,Du2025GSA,Yang2026,YanKX2026QB,REforQB2026Yang}. These studies suggest that dissipation is not always detrimental~\cite{PhysRevA.100.043833,Carrega_2020}; it can help control how energy is accumulated and stored. Still, selecting a long-lived mode does not necessarily lead to more extractable work, since the gain needed to amplify it also induces fluctuations.

The work quality of a quantum battery therefore depends on how coherent driving and gain-induced fluctuations accumulate during amplification~\cite{Extractable2022PRL}. A coherent drive produces a phase-defined response, so contributions from different times can reinforce or cancel one another. Incoherent gain noise, by contrast, consists of independent fluctuations that accumulate without the same temporal interference~\cite{Clerk2010}. This difference becomes important when multiple unstable modes contribute. With coherent coupling, two frequency-split unstable modes can interfere and hinder the buildup of extractable work. In contrast, dissipative coupling through a common reservoir can favor a single dominant mode and suppress such interference. However, whether such engineering can improve stored-work quality remains unclear.
\begin{figure}[b]
\centering
\includegraphics[width=1\linewidth]{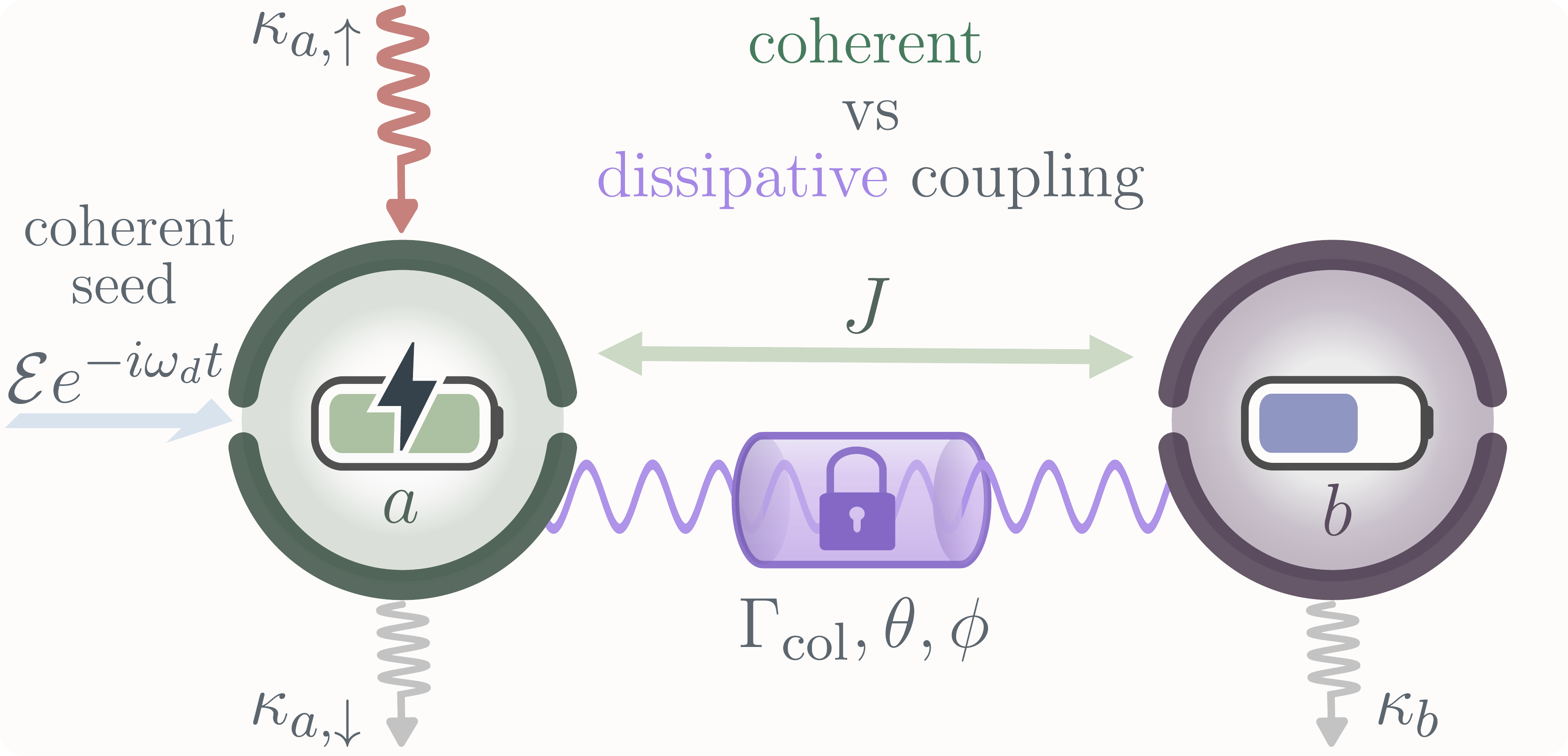}
\caption{Phase-locked quantum-battery charging.
Engineered dissipation selects an exponentially growing quasi-dark mode, suppressing coherent-mode beating and enhancing stored-work quality. \justifying}
\label{Schematic}
\end{figure}

In this Letter, we investigate how reservoir engineering reshapes the gain-assisted charging and the balance between extractable work and energy fluctuations. We compare coherent- and dissipative-coupling schemes at matched coupling strengths and instability growth rates, revealing distinct stability thresholds and charging dynamics. At finite target occupations, we compare several important figures of merit, such as the charging time, average ergotropy power, charging efficiency, and work-like SNR. For the coherent-coupling scheme, we further optimize the driving frequency independently for each target occupation. The frequency-optimized coherent scheme achieves higher ergotropy power and greater charging efficiency, while dissipative coupling yields a higher work-like SNR, indicating improved stored-work quality. To explain this SNR advantage, we derive an exact finite-time source–noise relation that links temporal response matching to the ratio of ergotropy to passive energy. At fixed target occupation, this ratio also determines the work-like SNR. Our results highlight the importance of mode selection for improving stored-work quality in the context of quantum-battery charging.



\emph{Model.---}We consider a two-mode bosonic system consisting of a charger mode $a$ and a storage mode $b$, with resonance frequencies $\omega_{a}$ and $\omega_{b}$, as shown in Fig.~\ref{Schematic}. The charger is coherently driven by an external field with amplitude $\mathcal{E}$ and frequency $\omega_d$. In a frame rotating at $\omega_d$, the system Hamiltonian reads ($\hbar=1$ hereafter)
\begin{equation}
H_0 = (\delta_d+\Delta)\,a^{\dagger}a
+(\delta_d-\Delta)\,b^{\dagger}b
+\mathcal{E}\,a^{\dagger}
+\mathcal{E}^{*}\,a,
\label{eq:H_main}
\end{equation}
where $a$ and $b$ ($a^\dag$ and $b^\dag$) are the annihilation (creation) operators of the charger and battery modes, respectively, $\Delta=(\omega_a-\omega_b)/2$ is charger--battery frequency mismatch, and $\delta_d=(\omega_a+\omega_b)/2-\omega_d$ is the drive detuning from their frequency midpoint. 
We compare two mechanisms for coupling the charger and the battery. For \emph{coherent} coupling, the interaction is described by $H_{\rm coh} = Ja^\dagger b+\mathrm{H.c.}$ and the total Hamiltonian of the two-mode setup is $H=H_0+H_{\rm coh}$, with $J$ being the coherent coupling rate. For \emph{dissipative} coupling, the two modes interact through a common engineered reservoir and no direct coherent interaction is present, so that $H=H_0$.


The Markovian dynamics of the setup is governed by
\begin{align}
\dot{\rho} = 
&-i[H,\rho]
+\kappa_{a,\downarrow}\mathcal{D}[a]\rho
+\kappa_{a,\uparrow}\mathcal{D}[a^\dagger]\rho
\nonumber\\
&+\kappa_b\mathcal{D}[b]\rho
+\Gamma_{\rm col}\mathcal{D}[z_{\theta,\phi}]\rho ,
\label{eq:master}
\end{align}
where $\mathcal{D}[L]\rho=L\rho L^\dagger-\frac12\{L^\dagger L,\rho\}$ and
\begin{equation}
z_{\theta,\phi} = \cos\theta\,a+e^{i\phi}\sin\theta\,b
\end{equation}
is a normalized collective jump operator. Here, \(\theta\) controls the relative coupling of the two modes to the common reservoir, while \(\phi\) determines their relative phase (we assume $\phi=0$ hereafter, without loss of generality). The collective channel contributes local dissipative terms proportional to $\Gamma_{\rm col}\cos^2\theta$ and $\Gamma_{\rm col}\sin^2\theta$, together with an off-diagonal dissipative coupling $g = \frac{\Gamma_{\rm col}}{2} e^{i\phi}\sin\theta\cos\theta$.
We set $\Gamma_{\rm col}=0$ for the coherent-coupling scheme and $J=0$ for the dissipative-coupling scheme. The rates $\kappa_{a,\downarrow}$ and $\kappa_b$ describe the local charger and battery losses, respectively, whereas $\kappa_{a,\uparrow}$ is the incoherent pump (gain) applied to the charger. The collective reservoir is passive, i.e. the active energy is supplied only by the coherent drive and incoherent pump.

In the following, we use $\kappa_0$ as the reference rate. Unless otherwise stated, frequencies are expressed in units of $\kappa_0$, time in units of $\kappa_0^{-1}$, and battery energies in units of $\omega_b$.
\begin{figure}[t]
\centering
\includegraphics[width=1\linewidth]{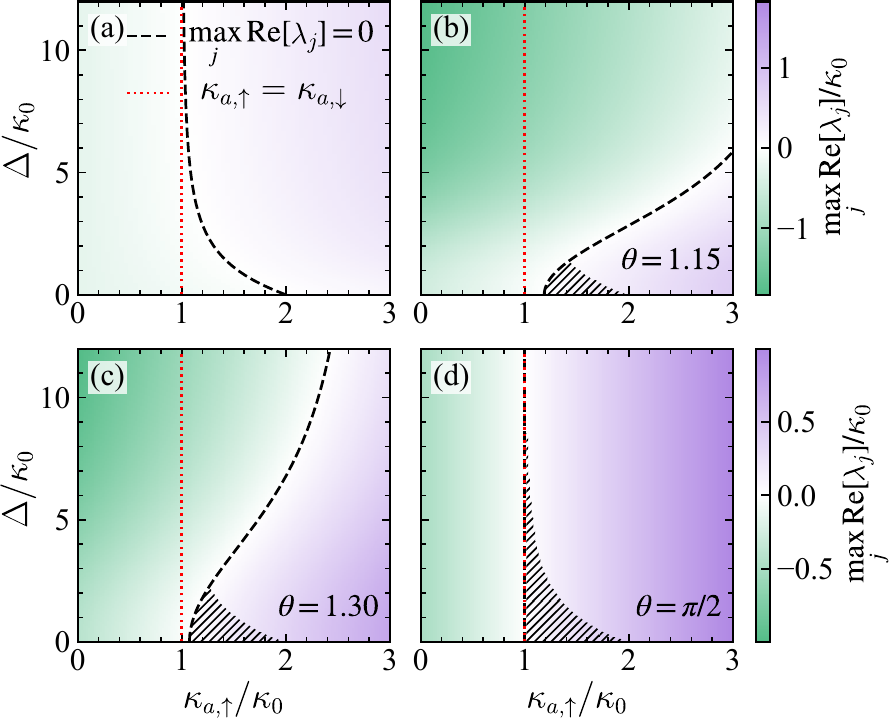}
\caption{Stability landscapes for the coherent- and dissipative-coupling schemes as functions of the charger--battery frequency mismatch $\Delta$ and the incoherent pump rate
$\kappa_{a,\uparrow}$. Panel (a) shows the coherent scheme with $J/\kappa_0=3.2$, while panels (b)--(d) show the dissipative scheme with $\Gamma_{\rm col}/\kappa_0=17.165$ and $\theta=1.15$, $1.30$, and $\pi/2$, respectively, corresponding to $g/\kappa_0=3.2$, $2.21$, and $0$. We set $\delta_d=0$ and $\kappa_{a,\downarrow}/\kappa_0=\kappa_b/\kappa_0=1$. The color scale shows $\max_j\mathrm{Re}[\lambda_j]/\kappa_0$. The black dashed curves mark the instability threshold, while the red dotted lines indicate $\kappa_{a,\uparrow}=\kappa_{a,\downarrow}$. Hatching regions indicate where the dissipative scheme is unstable while the coherent reference remains stable. 
\justifying}
\label{fig:stability_normalized}
\end{figure}

\emph{Stability.---}Before studying the charging dynamics, we first identify the stability conditions for the two coupling schemes. The linear dynamics governed by Eq.~\eqref{eq:master} becomes unstable when the corresponding drift matrix has an eigenvalue $\lambda_j$ with $\mathrm{Re}[\lambda_j]>0$ (see the Supplemental Material (SM)~\cite{SuppMat} for details). Thus, $\max_j\mathrm{Re}[\lambda_j]=0$ defines the marginal-stability boundary~\cite{ahmadi2025reservoir}. 

For the coherent-coupling scheme in the strong-hybridization regime, the on-resonance ($\Delta=0$) instability threshold is $\kappa_{a,\uparrow}^{\rm crit,coh} = \kappa_{a,\downarrow}+\kappa_b$~\cite{SuppMat}.
For the dissipative-coupling scheme, the corresponding on-resonance threshold is~\cite{SuppMat}
\(
\kappa_{a,\uparrow}^{\rm crit,dis}
= \kappa_{a,\downarrow}
+ \frac{\Gamma_{\rm col}\kappa_b\cos^2\theta}{\kappa_b
+ \Gamma_{\rm col}\sin^2\theta}.
\)
The latter reduces to $\kappa_{a,\uparrow}^{\rm crit,dis} \simeq \kappa_{a,\downarrow} + \kappa_b \cot^2{\theta}$ for $\Gamma_{\rm col}\sin^2{\theta}\gg\kappa_b$. 
As explained below, this reduced threshold originates from the emergence of a weakly damped quasi-dark collective mode, whose residual loss can be compensated by a relatively weak incoherent pump. Although the threshold approaches $\kappa_{a,\downarrow}$ as $\theta\rightarrow\pi/2$, the dissipative coupling $|g|\propto\sin\theta\cos\theta$ vanishes simultaneously at fixed $\Gamma_{\rm col}$. This limiting case therefore does not provide useful charging dynamics.

Figure~\ref{fig:stability_normalized} illustrates how the instability thresholds depend on the collective-mode angle $\theta$ and the charger--battery frequency mismatch $\Delta$ for a given parameter set.
For the coherent-coupling scheme [Fig.~\figpanel{fig:stability_normalized}{a}], the instability threshold is $\kappa_{a,\uparrow}^{\rm crit,coh}/\kappa_0=2$ when $\Delta=0$ and decreases toward the charger decay rate $\kappa_{a,\downarrow}$ as $\Delta$ increases, owing to the suppressed mode hybridization.
For dissipative coupling, increasing $\theta$ from $1.15$ to $1.3$ reduces the on-resonance threshold from $\kappa_{a,\uparrow}^{\rm crit,dis}/\kappa_0\simeq1.1872$ [Fig.~\figpanel{fig:stability_normalized}{b}] to $\simeq1.0725$ [Fig.~\figpanel{fig:stability_normalized}{c}], while its dependence on $\Delta$ also becomes weaker.
In the limiting case $\theta=\pi/2$ [Fig.~\figpanel{fig:stability_normalized}{d}], the collective reservoir acts solely on the battery and $\kappa_{a,\uparrow}^{\rm crit,dis}=\kappa_{a,\downarrow}$ becomes independent of $\Delta$.
However, the dissipative coupling $|g|=(\Gamma_{\rm col}/2)\sin\theta\cos\theta$ also vanishes in this limit, so the battery decouples from the charger. This motivates the choice of an intermediate operating point for the dissipative charging scheme (see SM~\cite{SuppMat} for details).
\begin{figure}[t]
\centering
\includegraphics[width=\columnwidth]{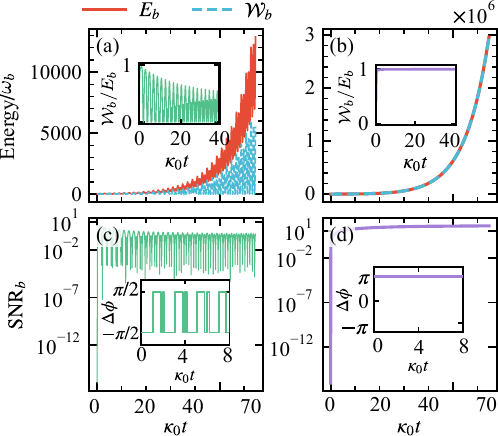}
\caption{Charging dynamics for coherent [(a) and (c)] and dissipative [(b) and (d)] coupling.
(a), (b) Battery energy $E_b$ (solid) and ergotropy
$\mathcal W_b$ (dashed), with the insets showing the ergotropy
fraction $\mathcal W_b/E_b$.
(c), (d) Work-like SNR $\mathcal{W}_b/\sigma_{E,b}$, with the insets showing the relative phase
$\Delta\phi=\arg\langle b\rangle-\arg\langle a\rangle$.
Both schemes have matched coupling strengths $J/\kappa_0=g/\kappa_0=3.2$ and maximal growth rates $\lambda_{\max}/\kappa_0=0.05$.
The coherent scheme uses $\kappa_{a,\uparrow}/\kappa_0=2.2$, while the dissipative scheme uses $\theta=1.15$, $\Gamma_{\rm col}/\kappa_0=17.165$, $\kappa_{a,\uparrow}/\kappa_0=1.3046$, and $\phi=0$.
Common parameters are $\kappa_{a,\downarrow}/\kappa_0=\kappa_b/\kappa_0=1$, $\mathcal{E}/\kappa_0=10$, and $\Delta=\delta_d=0$. 
\justifying}
\label{fig:matched_growth_dynamics}
\end{figure}

\emph{Charging dynamics.---}Since the dynamics is Gaussian, it is fully characterized by the first moments and covariance matrix, which obey a closed linear equation and a Lyapunov equation, respectively~\cite{Serafini2017QuantumCV}. In this case, the stored battery energy is given by~\cite{Campaioli2024Colloquium}
\(
E_b(t)=\langle H_b\rangle_t,
\)
with $H_b=\omega_b b^\dagger b$ and $\langle O\rangle_t=\mathrm{Tr}[O\rho(t)]$.
For the vacuum and thermal initial states considered below, the reduced battery state remains a displaced thermal state, whose ergotropy, defined as the maximum work extractable by a cyclic unitary, is given by~\cite{Allahverdyan2004,PhysRevLett.134.220402,2jtp-jpkn}
\(
\mathcal{W}_b(t)=\omega_b \left|\langle b\rangle_t \right|^2,
\)
while the passive fluctuation energy is
\(
E_b^{\rm pas}(t)=E_b(t)-\mathcal{W}_b(t).
\)
To assess the reliability of the stored energy in the presence of gain-induced fluctuations, we introduce the work-like SNR~\cite{AhmadiChiral2026}: 
\begin{equation}
\mathrm{SNR}_b(t)
=\frac{\mathcal W_b(t)}{\sigma_{E,b}(t)},
\label{eq:SNR_main}
\end{equation}
where 
$\sigma_{E,b}(t)
=\sqrt{\langle H_b^2\rangle_t-\langle H_b\rangle_t^2}$ is the standard deviation of the battery energy. For displaced thermal states, $\mathrm{SNR}_b$ can be expressed analytically in terms of the coherent and incoherent occupations, as derived in the SM~\cite{SuppMat}.

Figure~\ref{fig:matched_growth_dynamics} compares the charging dynamics of the two coupling schemes at matched coupling strengths, \(J/\kappa_0=g/\kappa_0=3.2\), assumed to be real without loss of generality, and matched maximal growth rate, \(\lambda_{\max}/\kappa_0=0.05\). Despite these matched conditions, the two schemes exhibit qualitatively different charging behaviors. For coherent coupling, frequency splitting produces two unstable hybrid modes with the same growth rate but opposite oscillation frequencies. Their interference leads to the persistent beating shown in Fig.~\figpanel{fig:matched_growth_dynamics}{a}. The ergotropy fraction $\mathcal{W}_b/E_b$ also exhibits pronounced oscillations, since the interference between the two modes periodically suppresses the coherent battery amplitude. Dissipative coupling instead produces a linewidth splitting between a strongly damped bright mode and a weakly damped quasi-dark mode. Only the latter is unstable at the chosen operating point, so that the dynamics is dominated by a single exponentially growing mode after a short transient. Consequently, the beating is suppressed and $\mathcal{W}_b$ closely follows $E_b$, with their ratio approaching unity, as shown in Fig.~\figpanel{fig:matched_growth_dynamics}{b}.

The two coupling schemes also differ in their work-like SNR. The coherent scheme exhibits pronounced oscillations, whereas the dissipative scheme yields a monotonically increasing \(\mathrm{SNR}_b\) that reaches substantially larger values at long times [Figs.~\figpanel{fig:matched_growth_dynamics}{c} and \figpanel{fig:matched_growth_dynamics}{d}]. The insets in Figs.~\figpanel{fig:matched_growth_dynamics}{c} and ~\figpanel{fig:matched_growth_dynamics}{d} further show the relative phase $\Delta\phi(t)=\arg\langle b(t)\rangle-
\arg\langle a(t)\rangle$ between the charger and the battery. For coherent coupling, $\Delta\phi$ switches between $\pm\pi/2$, reflecting the interference between the two oscillatory unstable modes. By contrast, dissipative
coupling maintains a constant relative phase $\Delta\phi=\pi$, indicating \emph{phase locking} between the charger and battery. The coherent drive provides the phase reference for the selected quasi-dark mode, which is subsequently amplified by the incoherent pump.
%
\begin{figure}[t]
\centering
\includegraphics[width=\columnwidth]{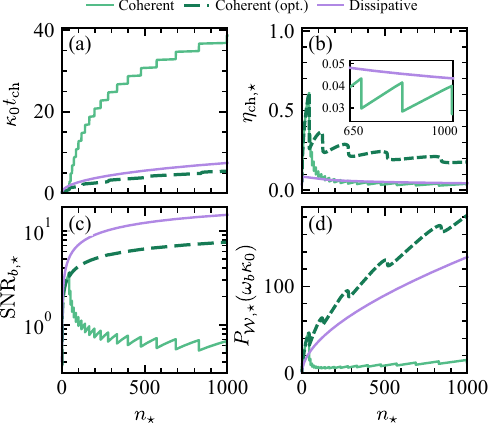}
\caption{Finite-capacity charging performance of the coherent- and dissipative-coupling schemes over
$0.2\leq n_\star\leq1000$.
(a) First-passage charging time $t_{\rm ch}$.
(b) Charging efficiency $\eta_{{\rm ch},\star}$.
(c) Work-like SNR ${\rm SNR}_{b,\star}=\mathcal W_{b,\star}/\sigma_{E,b,\star}$.
(d) Average ergotropy power
$P_{\mathcal W,\star}$. 
The coherent scheme uses resonant driving (green solid) with $\delta_d=0$, or frequency-optimized driving (dark-green dashed) with \(\omega_d\) optimized at each \(n_\star\) to minimize \(t_{\rm ch}\). The dissipative scheme uses resonant driving (purple solid). The inset in (b) magnifies the large-$n_\star$ regime to highlight the difference between the dissipative and resonantly-driven coherent schemes. All other parameters are the same as those in Fig.~\ref{fig:matched_growth_dynamics}. \justifying}
\label{fig:finite_capacity_comparison}
\end{figure}

%

\textit{Finite-capacity performance.---}The linear model studied above has no intrinsic capacity and predicts unbounded energy growth. In practice, the accessible battery occupation may be limited by a finite storage capacity or by saturation of the gain medium. We therefore evaluate the charging performance at the first-passage charging time \(t_{\rm ch}(n_\star) = \inf\{t\geq 0:\langle b^\dagger b\rangle(t)\geq n_\star\}\), i.e., the earliest time at which the battery occupation reaches a prescribed target \(n_\star\). A complementary realization with a finite, saturable gain medium is discussed in the End Matter. At the first-passage time $t_{\rm ch}(n_\star)$, we evaluate the charging schemes with the work-like SNR
\(
\mathrm{SNR}_{b,\star} = \mathrm{SNR}_b[t_{\rm ch}(n_\star)]
\)
and the average ergotropy power 
\begin{equation}
P_{\mathcal{W},\star}
=
\frac{\mathcal{W}_b[t_{\rm ch}(n_\star)]}
{t_{\rm ch}(n_\star)}.
\label{eq:P_W}
\end{equation}
For the finite-capacity case, we also introduce the charging efficiency
\begin{equation}
\eta_{{\rm ch},\star} = \frac{\mathcal W_b[t_{\rm ch}(n_\star)]}
{E_{\rm act}[t_{\rm ch}(n_\star)]},
\label{eq:efficiency}
\end{equation}
where $E_{\rm act}$ is the active-energy cost associated with the coherent drive and incoherent gain, as detailed in the End Matter.

Figure~\ref{fig:finite_capacity_comparison} compares these quantities over the target-occupation range \(0.2\leq n_\star\leq1000\). For both schemes, we first consider resonant driving with $\omega_d=\omega_a=\omega_b$ (i.e., $\Delta=\delta_d=0$). At small target occupations, the coherent-coupling scheme reaches the target earlier, as shown in Fig.~\figpanel{fig:finite_capacity_comparison}{a}. As $n_\star$ increases, however, the oscillatory charging dynamics in Fig.~\figpanel{fig:matched_growth_dynamics}{a} produces a series of local maxima in the battery occupation. Once a target exceeds one such maximum, it can only be reached during a later oscillation cycle, leading to
a step-like increase in $t_{\rm ch}$. By contrast, the phase-locked dissipative-coupling scheme exhibits a smooth increase in $t_{\rm ch}$ and
eventually charges faster than the resonantly-driven coherent scheme at large $n_\star$.

The charging efficiency shows a different trend, as shown in Fig.~\figpanel{fig:finite_capacity_comparison}{b}. Here the coherent-drive contribution to $E_{\rm act}$ is treated as a \emph{signed} energy transfer, so that the energy returned from the system to the coherent drive reduces the accumulated input energy. Under this convention, the coherent scheme achieves a higher charging efficiency at small $n_\star$, partly because its oscillatory dynamics allows energy to flow back toward the drive. At larger $n_\star$, however, the beating-induced charging delays substantially increase the accumulated energy input from the incoherent pump and lead to a sharp drop in efficiency. By contrast, the dissipative scheme avoids these delays through phase-locked amplification and exhibits a smoother efficiency profile, eventually outperforming the resonantly-driven coherent scheme. This efficiency ordering depends on the energy-accounting convention, as discussed in the End Matter.

The advantage of dissipative coupling is particularly pronounced in the quality of the stored energy [Fig.~\figpanel{fig:finite_capacity_comparison}{c}]. 
Its work-like SNR is substantially higher than that of the coherent-coupling scheme under resonant driving.
Moreover, the average ergotropy power in Fig.~\figpanel{fig:finite_capacity_comparison}{d} exhibits a crossover between the two resonantly driven schemes. While coherent coupling enables higher power at low target occupations,
the beating-induced charging delays sharply reduce its power as \(n_\star\) increases. In contrast, dissipative coupling avoids these delays and eventually enables higher power.

However, coherent coupling splits the hybrid-mode frequencies, so that the resonant drive excites both modes with comparable amplitudes, leading to interference that can substantially delay charging. To mitigate this effect, we optimize the coherent driving frequency $\omega_d$ for each $n_\star$ independently to minimize the charging time,
while keeping the drive amplitude, coupling strengths, and maximal growth rate fixed. For the dissipative scheme, resonant driving is optimal for all target occupations considered here (see SM~\cite{SuppMat}). As shown by the dark-green dashed curves in Fig.~\ref{fig:finite_capacity_comparison}, the frequency optimization largely eliminates the coherent charging delays. As a result, the optimized coherent-coupling scheme charges faster and achieves higher charging efficiency than the dissipative scheme throughout the considered target-occupation range. Nevertheless, dissipative coupling retains a significantly larger work-like SNR, as shown in Fig.~\figpanel{fig:finite_capacity_comparison}{c}. This SNR advantage persists over a broad range of coherent (\(J\)) and dissipative (\(g\)) coupling strengths at fixed pump rates (see SM \cite{SuppMat} for details).



\emph{Temporal response matching.---}To understand the work-like SNR advantage of the dissipative-coupling scheme, we first introduce the charger-to-battery response
$h_{ba}(u)=\bm e_b^\dagger e^{Mu}\bm e_a$, where $M$ is the drift matrix~\cite{SuppMat}.
For vacuum initial conditions, the coherent drive and incoherent gain noise enter through the same charger mode, but their contributions accumulate differently. The coherent amplitudes can interfere destructively, whereas the noise contributions accumulate independently.
This yields the exact finite-time relation (see SM~\cite{SuppMat} for derivations)
\begin{equation}
\begin{aligned}
R_b(t)&\equiv\frac{\mathcal W_b(t)}{E_b^{\rm pas}(t)}
=\frac{|\mathcal E|^2t}{\kappa_{a,\uparrow}}
\mathcal M_b(t),\\
\mathcal M_b(t)&=
\frac{\left|\int_0^t h_{ba}(u)\,du\right|^2}
{t\int_0^t|h_{ba}(u)|^2\,du}\leq1.
\end{aligned}
\label{eq:source_noise_matching}
\end{equation}
%
The dimensionless factor $\mathcal M_b$ quantifies the \emph{temporal response matching}. The Cauchy--Schwarz inequality gives $0\leq\mathcal M_b\leq1$ whenever the denominator is nonzero, with equality at unity only when the complex response is constant over the time interval. Thus, variations in either the response amplitude or phase reduce $\mathcal M_b$. 


At $n_\star=1000$, $\mathcal M_b$ is approximately $4.9\times10^{-4}$, $0.5$, and $0.98$ for the resonant coherent, optimized coherent, and dissipative schemes, respectively, implying that the coherent-drive response undergoes much less temporal cancellation in the dissipative scheme. The ratio $R_b$ also depends on the gain rate and charging time through the prefactor in Eq.~\eqref{eq:source_noise_matching}. At a fixed target occupation, the work-like SNR increases monotonically with \(R_b\), as shown in the SM~\cite{SuppMat}. The stronger temporal response matching of the dissipative scheme thus provides a physical explanation for its enhanced work-like SNR.

\textit{Conclusions.---}Our results show that the performance of gain-assisted quantum charging depends on both the growth rate and the structure of the amplified modes. While coherent coupling gives rise to two interfering unstable modes, dissipative reservoir engineering
selects a single quasi-dark unstable mode, enabling exponential charging with a near-unit ergotropy fraction. 
The finite-time source--noise relation links temporal response matching to the ratio of ergotropy to passive energy and, at fixed battery occupation, to the work-like SNR. This explains why the dissipative-coupling scheme retains a work-quality advantage even when the frequency-optimized coherent-coupling scheme achieves higher ergotropy power. 


An important open question is whether nonlinear gain saturation can stabilize a charged state while preserving the high ergotropy generated by the transient
amplification~\cite{Barra2019}.
Extending mode-selective amplification to multimode systems may also enable phase-coherent energy storage across distributed batteries through engineered dark subspaces~\cite{Diehl2008}. Combining this approach with nonlinear reservoir engineering could further extend mode-selective charging to non-Gaussian battery states~\cite{Leghtas2015}, opening new possibilities for work storage beyond coherent displacement and Gaussian squeezing.


\acknowledgments{\emph{Acknowledgments}---B.A. and P.H. acknowledge support from IRA Programme (project no. FENG.02.01-IP.05-0006/23) financed by the FENG program 2021-2027, Priority FENG.02, Measure FENG.02.01., with the support of the FNP. L.D. acknowledges financial support by the Knut och Alice Wallenberg stiftelse through project grant no. 2022.0090. A.H.A.M. acknowledges support from the National Science Centre, Poland (SONATINA-9, grant agreement no. UMO-2025/56/C/ST2/00368 entitled “Modularne kwantowe urządzenia termiczne: integrowanie termicznych funkcjonalności”). S.B. acknowledges financial support from the Natural Sciences and Engineering Research Council of Canada (NSERC) through a Discovery Grant and from the University of Calgary. A.M. acknowledges support from the QEC4QEA project through EuroHPC JU (Grant No. 101194322) and the Polish NCBR (Grant No. DWM/EuroHPC/2023/429/2025).}

\textit{Data and code availability}.---
The numerical data and figures presented in the main text and Supplemental Material can be reproduced using the simulation code, plotting routines, and parameter files available at \cite{PhaseLockingCode}.

\bibliography{References}

\section*{End Matter}

\textit{Active-energy accounting.---}To calculate the charging-efficiency behavior in Fig.~\figpanel{fig:finite_capacity_comparison}{b}, we decompose the active-energy cost into contributions from the incoherent pump (gain) and coherent drive,
\begin{equation}
E_{\rm act}(t) = E_{\rm gain}(t) + E_{\rm drive}(t).
\label{eq:Eact}
\end{equation}
The incoherent-gain contribution is
\begin{equation}
E_{\rm gain}(t) = \omega_a\kappa_{a,\uparrow}\int_0^t\left(\langle n_a(t')\rangle+1\right)dt',
\label{eq:Egain}
\end{equation}
while the coherent-drive contribution is
\begin{equation}
E_{\rm drive}(t) = \int_0^t P_{\rm drive}(t')\,dt',
\label{eq:Edrive}
\end{equation}
with
\(P_{\rm drive}(t) = -2\omega_a\,{\rm Im}\left[\mathcal{E}^*\langle a(t)\rangle\right]\). The coherent-drive contribution is treated as a \emph{signed} energy transfer. A positive $P_{\rm drive}$ corresponds to energy injected into the system, while a negative $P_{\rm drive}$ indicates energy flowing from the system back toward the coherent source.

\begin{figure}[t]
\centering
\includegraphics[width=\columnwidth]{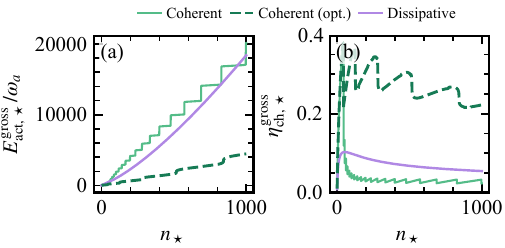}
\caption{Gross-input charging performance as a function of the target occupation
$n_\star$.
(a) Gross active-energy input $E_{{\rm act},\star}^{\rm gross}
=E_{{\rm gain},\star}+E_{{\rm in},\star}^{\rm coh}$.
(b) Corresponding charging efficiency
$\eta_{{\rm ch},\star}^{\rm gross} = \mathcal W_b(t_{\rm ch})/E_{{\rm act},\star}^{\rm gross}$.
The coherent scheme uses resonant driving (green solid) or frequency-optimized driving (dark-green dashed), while the dissipative scheme uses resonant driving (purple solid). We assume $|a_{\rm in}|^2=|\mathcal E|^2/\kappa_{\rm ext}$ and $\kappa_{\rm ext}=\kappa_{a,\downarrow}=\kappa_0$.
All other parameters are the same as those in
Fig.~\ref{fig:finite_capacity_comparison}. \justifying}
\label{fig:Fig5_gross_input}
\end{figure}

\textit{Gross-input efficiency.---}Under the signed-input convention of Eq.~\eqref{eq:Edrive}, energy returned from the system to the coherent source reduces the net input. However, such returned energy may not be recoverable in practice. We therefore consider a gross-input convention that counts the total incident coherent-field energy,
\begin{equation}
E_{\rm in}^{\rm coh}(t) = \omega_d\int_0^t |a_{\rm in}(t')|^2 dt'
\end{equation}
with \(|a_{\rm in}|^2 = |\mathcal E|^2/\kappa_{\rm ext}\).
Accordingly, we define the \emph{gross} active-energy cost
\begin{equation}
E_{\rm act}^{\rm gross} = E_{\rm gain}+E_{\rm in}^{\rm coh}
\label{eq:E_act_gross}
\end{equation}
and the corresponding charging efficiency
\begin{equation}
\eta_{\rm ch,\star}^{\rm gross} = \frac{\mathcal W_b[t_{\rm ch}(n_\star)]}
{E_{\rm act}^{\rm gross}[t_{\rm ch}(n_\star)]}.
\label{eq:gross_efficiency}
\end{equation}
Figure~\ref{fig:Fig5_gross_input} compares the gross-input charging performance of the coherent- and dissipative-coupling schemes, with
$\kappa_{\rm ext}=\kappa_{a,\downarrow}=\kappa_0$. As shown in the main text, under resonant driving, the dissipative-coupling scheme reaches large target occupations sooner than the coherent-coupling scheme [Fig.~\figpanel{fig:finite_capacity_comparison}{a}], and thereby requires less coherent-field input at a fixed drive amplitude. Consequently, at large target occupations, it requires less gross active-energy cost [Fig.~\figpanel{fig:Fig5_gross_input}{a}] and achieves higher gross-input efficiency [Fig.~\figpanel{fig:Fig5_gross_input}{b}], reversing the efficiency ordering of the resonantly-driven schemes under the signed-input convention [Fig.~\figpanel{fig:finite_capacity_comparison}{b}].
However, optimizing the coherent driving frequency largely avoids the beating-induced charging delays and substantially reduces the gross active-energy cost of the coherent-coupling scheme. As a result, the optimized coherent scheme achieves a higher gross-input efficiency than the dissipative-coupling scheme at large target occupations. Thus, the efficiency ordering depends on both the driving protocol and the energy-accounting convention.
%

\begin{figure}[t]
\centering
\includegraphics[width=\columnwidth]{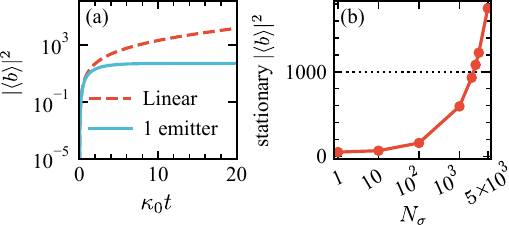}
\caption{Finite-gain saturation and emitter-number scaling.
(a) Time evolution of the coherent battery occupation
$|\langle b\rangle|^2$ for the linear-gain model (dashed)
and the saturable single-emitter model (solid).
(b) Stationary coherent battery occupation
$|\langle b\rangle|^2$ as a function of the emitter
number $N_\sigma$.
The individual emitter-charger coupling is scaled as
$g_\sigma^{(N_\sigma)}=g_\sigma^{(1)}/\sqrt{N_\sigma}$,
keeping the small-signal growth rate fixed as
$N_\sigma$ varies.
The horizontal dotted line in panel (b) indicates $|\langle b\rangle|_{\rm ss}^2=1000$, which is exceeded for
$N_\sigma\gtrsim 2.25\times10^3$.
Both panels use
$(\kappa_{a,\downarrow},\kappa_b,\Gamma_{\mathrm{col}},
\mathcal{E},P_\sigma,\gamma_\sigma,g_\sigma^{(1)})
/\kappa_0=(1,1,17.165,10,18,2,2.86258)$,
$\theta=1.15$, and $\delta_d=\Delta=\Delta_\sigma
=\gamma_{\sigma,\phi}=0$.
The corresponding small-signal maximal growth
rate is $\lambda_{\max}/\kappa_0=0.05$ (corresponding to $\kappa_{a,\uparrow}/\kappa_0=1.3046$ in the linear-gain model).
In both panels, the pumped-emitter model is treated within the Maxwell–Bloch mean-field approximation. 
\justifying}
\label{fig:saturation}
\end{figure}

\textit{Finite-gain saturation.}---The linear gain channel
$\kappa_{a,\uparrow}\mathcal D[a^\dagger]$ neglects gain depletion and therefore allows unbounded exponential growth in the unstable regime. To account for gain saturation, we replace this channel with $N_\sigma$ incoherently pumped two-level emitters coupled to the charger. This model provides a microscopic realization of gain saturation, which completes the finite-capacity analysis in the main text. The resulting master equation is
\begin{align}
\dot{\rho}=&-i[H_N,\rho]
+\kappa_{a,\downarrow}\mathcal D[a]\rho
+\kappa_b\mathcal D[b]\rho \notag\\
&+\Gamma_{\rm col}\mathcal D[z_{\theta,\phi}]\rho
+\sum_{j=1}^{N_\sigma}
\left(
P_\sigma\mathcal D[\sigma_+^{(j)}]
+\gamma_\sigma\mathcal D[\sigma_-^{(j)}]
\right.\notag\\
&\left.\qquad\qquad
+\frac{\gamma_{\sigma,\phi}}{2}
\mathcal D[\sigma_z^{(j)}]
\right)\rho,
\end{align}
where
\begin{align}
H_N={}&H_0+
\Delta_\sigma\sum_{j=1}^{N_\sigma}
\sigma_+^{(j)}\sigma_-^{(j)}
\notag\\
&+g_\sigma^{(N_\sigma)}
\sum_{j=1}^{N_\sigma}
\left(a^\dagger\sigma_-^{(j)}
+a\sigma_+^{(j)}\right).
\end{align}
Here $P_\sigma$, $\gamma_\sigma$, and $\gamma_{\sigma,\phi}$ denote the pumping, relaxation, and pure-dephasing rates of the emitters, respectively. For $P_\sigma>\gamma_\sigma$, the emitters are population inverted in their uncoupled steady state. When coupled to the charger, they can provide approximately linear gain at low charger occupations. As the charger population grows, stimulated emission depletes the inversion and therefore reduces the effective gain until it balances the losses.
The single-emitter calculation in Fig.~\figpanel{fig:saturation}{a} shows that this saturation mechanism suppresses the exponential growth, yielding a finite stationary coherent occupation in the battery mode.

The attainable battery occupation can be increased
by using more emitters. To compare different emitter numbers at a fixed collective coupling strength, we scale the individual emitter-charger coupling as
$g_\sigma^{(N_\sigma)} = g_\sigma^{(1)}/\sqrt{N_\sigma}$.
As shown in Fig.~\figpanel{fig:saturation}{b}, increasing $N_\sigma$
substantially enhances the steady-state coherent battery occupation. More specifically, for the parameters considered,
the Maxwell--Bloch calculation~\cite{Cartar2017} predicts that reaching
the stationary coherent occupation
$|\langle b\rangle|_{\rm ss}^2=1000$
requires approximately
$N_\sigma=2.25\times10^3$ emitters.


The above finite-gain analysis focuses on the dissipative-coupling scheme. Extending it to systems with both coherent and dissipative couplings, as considered in
Ref.~\cite{PhysRevLett.132.210402}, would be a natural next step. These results demonstrate that gain saturation can arrest the exponential growth while allowing substantial
energy storage in the battery. Further details are provided in SM~\cite{SuppMat}.

\emph{Gauge-invariant origin.---}The distinction between the coherent- and dissipative-coupling schemes is independent of the phase convention chosen for the two modes. Although the phases of the drift-matrix off-diagonal elements are gauge dependent, their product is invariant. This product is negative for coherent coupling and positive for dissipative coupling, giving rise to frequency and linewidth splitting, respectively~\cite{SuppMat}.

As discussed in the main text, the linewidth splitting in the dissipative-coupling scheme can be understood in terms of bright and dark collective modes. The common reservoir directly damps the bright superposition $z_{\theta,\phi}$, while the orthogonal dark superposition decouples from this reservoir through destructive interference. Local losses give this dark mode a finite residual linewidth, while a finite detuning can further mix it with the bright mode. Once the incoherent gain overcomes the effective damping of the quasi-dark mode, this mode becomes unstable and dominates the long-time dynamics. This explains both the reduced instability threshold and the phase-locked single-mode amplification observed above. 

This mechanism is closely related to the zero-linewidth condition of bound states in the continuum (BICs) in dissipatively coupled systems~\cite{PhysRevLett.125.147202}. This dark-mode condition is distinct from an exceptional point, which requires the coalescence of both eigenvalues and eigenvectors. Approaching the dark-mode limit further suppresses the instability threshold while reducing the effective charger--battery coupling, thereby limiting energy transfer to the battery.

\clearpage
\onecolumngrid


\setcounter{section}{0}
\setcounter{subsection}{0}
\setcounter{equation}{0}
\setcounter{figure}{0}
\setcounter{table}{0}

\renewcommand{\thesection}{\arabic{section}}
\renewcommand{\thesubsection}{\Alph{subsection}}
\renewcommand{\theequation}{S\arabic{equation}}
\renewcommand{\thefigure}{S\arabic{figure}}
\renewcommand{\thetable}{S\arabic{table}}

\makeatletter
\renewcommand{\p@subsection}{\thesection}
\makeatother

\newcommand{\suppnote}[1]{%
    \refstepcounter{section}%
    \setcounter{subsection}{0}%
    \section*{Note \arabic{section}. #1}%
    \addcontentsline{toc}{section}{Supplementary Note \arabic{section}. #1}%
}

\newcommand{\suppsubsection}[1]{%
    \refstepcounter{subsection}%
    \subsection*{\Alph{subsection}. #1}%
    \addcontentsline{toc}{subsection}{\Alph{subsection}. #1}%
}

\section*{Supplementary Materials for ``Phase-Locked Quantum Battery Charging via Dissipative Mode Selection''}

The Supplemental Material provides detailed derivations of the stability thresholds, quasi-dark-mode structure, and Gaussian equations governing the charging dynamics. We further examine finite-gain saturation, phase locking, and the finite-time source–noise relation underlying the stored-work quality.

\suppnote{Linear stability}
\label{app:exceptional_points}

To derive the instability thresholds discussed in the main text, we first consider the linear dynamics of the charger and battery modes, which are described by the annihilation (creation) operators $a$ and $b$ ($a^\dag$ and $b^\dag$), respectively. We then identify the physical origin of the reduced gain threshold in the dissipative-coupling scheme.

For the two-mode bosonic system introduced in the main text, we define the vector of first moments
\begin{equation}
\bm v(t)=\big(\langle a(t)\rangle,\langle b(t)\rangle\big)^{\mathsf T}.
\end{equation}
For bare-mode frequencies $\omega_a$ and $\omega_b$ and driving frequency $\omega_d$, the first-moment dynamics is governed by
\begin{equation}
\dot{\bm v}=H_r\bm v+\bm f,
\qquad
H_r=
\begin{pmatrix}
-\gamma_a-i(\delta_d+\Delta) & M_{ab}\\
M_{ba} & -\gamma_b-i(\delta_d-\Delta)
\end{pmatrix}.
\label{eq:Hr_affine}
\end{equation}
Here, $\Delta=(\omega_a-\omega_b)/2$ represents the charger--battery frequency mismatch, while $\delta_d=(\omega_a+\omega_b)/2-\omega_d$ denotes the drive detuning from the midpoint of the bare-mode frequencies. Since $\delta_d$ enters the drift matrix as a uniform shift $-i\delta_d\mathbb I$, varying the driving frequency does not change the eigenvalue real parts (and hence the instability growth rates). The quantities $\gamma_a$ and $\gamma_b$ are the \emph{effective decay rates}, which can be negative in the presence of gain, and $M_{ab}$ and $M_{ba}$ are the off-diagonal elements of the drift matrix $H_r$. The coherent drive enters through $\bm f= (-i\mathcal{E}, 0)^T$. The explicit coefficients for the coherent- and dissipative-coupling schemes are given below. 

Defining
\begin{equation}
\bar\gamma=\frac{\gamma_a+\gamma_b}{2},
\qquad
\Delta\gamma=\frac{\gamma_a-\gamma_b}{2},
\qquad
\Pi=M_{ab}M_{ba},
\end{equation}
the eigenvalues of the drift matrix are given by $\det(\lambda I-H_r)=0$, i.e.,
\begin{equation}
 \lambda_{\pm}
 =
 -\bar\gamma-i\delta_d
 \pm
 \sqrt{(\Delta\gamma+i\Delta)^2+\Pi}.
\label{eq:splus_general}
\end{equation}
The system is unstable when $\max_j\operatorname{Re}[\lambda_j]>0$. For real \(\Pi\) and \(\bar\gamma>0\), this condition is equivalent to
\begin{equation}
\Pi>
\gamma_a\gamma_b
\left[
1+
\frac{4\Delta^2}{(\gamma_a+\gamma_b)^2}
\right].
\label{eq:instab_unified}
\end{equation}
This criterion applies directly to the dissipative-coupling scheme. For coherent coupling on resonance, however, the two eigenvalues have the same real part $-\bar\gamma$ in the strong-hybridization regime. The instability threshold therefore occurs at $\bar\gamma=0$, where Eq.~\eqref{eq:instab_unified} is not applicable and Eq.~\eqref{eq:splus_general} should be used instead.

\suppsubsection{Coherent and dissipative thresholds}

We now apply the general stability analysis above to the coherent- and dissipative-coupling schemes introduced in the main text. We first consider the coherent-coupling scheme, for which the interaction Hamiltonian is \(H_{\rm coh}=Ja^\dagger b+J^* b^\dagger a\). The corresponding drift matrix is
\begin{equation}
H_r^{\rm coh}
=
\begin{pmatrix}
-\gamma_a^{\rm coh}-i(\delta_d+\Delta) & -iJ\\
-iJ^* & -\gamma_b^{\rm coh}-i(\delta_d-\Delta)
\end{pmatrix},
\qquad
\gamma_a^{\rm coh}=\frac{\kappa_{a,\downarrow}-\kappa_{a,\uparrow}}{2},
\qquad
\gamma_b^{\rm coh}=\frac{\kappa_b}{2}.
\label{eq:coherent_complex_drift}
\end{equation}
The gauge-invariant product of the off-diagonal elements is
\begin{equation}
\Pi_{\rm coh}=-|J|^2.
\label{eq:bc_coherent}
\end{equation}
In the strong-hybridization regime $|J|>|\Delta\gamma^{\rm coh}|=|\gamma_a^{\rm coh}-\gamma_b^{\rm coh}|/2$ and for resonant bare modes ($\Delta=0$), the two eigenvalues have the same real part, $\rm{Re}[\lambda_{\pm}]=-\bar\gamma$. The instability threshold is therefore determined by $\bar\gamma=0$, yielding
\begin{equation}
\kappa_{a,\uparrow}^{\rm crit,coh}
=
\kappa_{a,\downarrow}+\kappa_b.
\label{eq:coherent_strong_coupling_threshold}
\end{equation}
For \(\kappa_{a,\downarrow}=\kappa_b=\kappa_0\), this gives \(\kappa_{a,\uparrow}^{\rm crit,coh}/\kappa_0=2\).

We next consider dissipative coupling mediated by the engineered common reservoir. As introduced in the main text, the collective jump operator of the two modes is~\cite{ReservoirEngineering2014,ReservoirEngineering2015}
\begin{equation}
z_{\theta,\phi}
=
\cos\theta\,a
+
e^{i\phi}\sin\theta\,b,
\end{equation}
where $\theta$ controls
the relative coupling of the two modes to the common
reservoir, while $\phi$ determines their relative phase. The resulting drift matrix takes the form
\begin{equation}
H_r^{\rm diss}
=
\begin{pmatrix}
-\gamma_a-i(\delta_d+\Delta) & -g\\
-g^* & -\gamma_b-i(\delta_d-\Delta)
\end{pmatrix},
\label{eq:H2_diss_normalized}
\end{equation}
where
\begin{equation}
g
=
\frac{\Gamma_{\rm col}}{2}
 e^{i\phi}\sin\theta\cos\theta,
\qquad
\gamma_a
=
\frac{\kappa_{a,\downarrow}-\kappa_{a,\uparrow}+\Gamma_{\rm col}\cos^2\theta}{2},
\qquad
\gamma_b
=
\frac{\kappa_b+\Gamma_{\rm col}\sin^2\theta}{2}.
\label{eq:gamma_ab_normalized}
\end{equation}
Different from the coherent-coupling scheme, the off-diagonal product is positive in this case:
\begin{equation}
\Pi_{\rm diss}=|g|^2\geq0.
\end{equation}
Substituting these coefficients into the general instability condition [cf. Eq.~\eqref{eq:instab_unified}], we obtain the threshold for the dissipative-coupling scheme. In particular, the on-resonance instability boundary is determined by $\Pi_{\rm diss}=\gamma_a^{\rm diss}\gamma_b^{\rm diss}$, yielding the simplified critical pump rate
\begin{equation}
\kappa_{a,\uparrow}^{\rm crit,diss}
=
\kappa_{a,\downarrow}
+
\frac{\Gamma_{\rm col}\kappa_b\cos^2\theta}
{\kappa_b+\Gamma_{\rm col}\sin^2\theta}.
\label{eq:kup_crit_diss_normalized}
\end{equation}
Unlike the coherent-coupling threshold in Eq.~\eqref{eq:coherent_strong_coupling_threshold}, this threshold depends explicitly on the collective dissipation rate $\Gamma_{\rm col}$ and the mode-composition angle $\theta$. As we explain in the following subsection, the reduced threshold originates from the formation of a weakly damped quasi-dark mode.

\begin{figure}[t]
\centering
\includegraphics[width=0.6\textwidth]{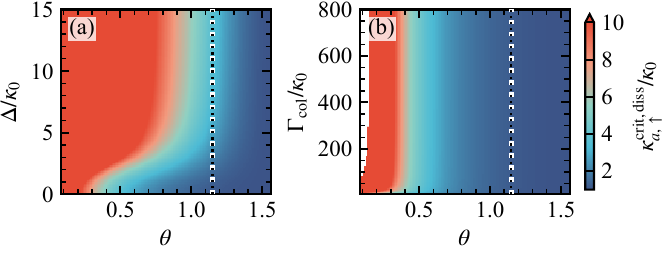}
\caption{Critical incoherent pump rate
$\kappa_{a,\uparrow}^{\rm crit,diss}/\kappa_0$
for the dissipatively coupled model ($J=0$), with
$\kappa_{a,\downarrow}/\kappa_0=\kappa_b/\kappa_0=1$. The critical pump is determined by the first stability boundary, $\max_j\operatorname{Re}[\lambda_j]=0$. (a) Dependence on $\theta$ and $\Delta/\kappa_0$ at fixed $\Gamma_{\rm col}/\kappa_0=17.165$, computed from the full drift matrix in
Eq.~\eqref{eq:H2_diss_normalized}.
(b) Dependence on $\theta$ and $\Gamma_{\rm col}/\kappa_0$ at bare-mode resonance ($\Delta=0$). Vertical dotted lines mark the reference operating
point $\theta=1.15$. The color scale indicates $\kappa_{a,\uparrow}^{\rm crit,diss}/\kappa_0$, with values above its upper limit shown in the saturated color. \justifying}
\label{fig:normalized_threshold_scans}
\end{figure}

Figure~\ref{fig:normalized_threshold_scans} shows the dependence of the dissipative instability threshold $\kappa_{a,\uparrow}^{\rm crit,diss}$ on the mode-composition angle \(\theta\), charger-battery frequency mismatch \(\Delta\), and collective decay rate \(\Gamma_{\rm col}\). In Fig.~\figpanel{fig:normalized_threshold_scans}{a}, the threshold is lowest at resonance and generally increases with \(|\Delta|\), while increasing \(\theta\) reduces the required pump rate. Figure~\figpanel{fig:normalized_threshold_scans}{b} shows that, at fixed \(0<\theta<\pi/2\), increasing \(\Gamma_{\rm col}\) raises the critical pump rate toward its strong-dissipation limit \(\kappa_{a,\downarrow}+\kappa_b\cot^2\theta\), in agreement with Eq.~\eqref{eq:kup_crit_diss_normalized}.

\suppsubsection{Quasi-dark mode, bound states in the continuum, and exceptional points}
\label{sec:bic_low_threshold}

We now explain how the common reservoir lowers the instability
threshold and why this does not necessarily improve battery
charging. The key is that the reservoir does not damp the
charger and battery independently. Instead, it couples to
a particular superposition of their fields. This gives rise
to two orthogonal collective modes: the bright mode
\begin{equation}
c_{\rm B}
=
\cos\theta\,a
+
e^{i\phi}\sin\theta\,b,
\end{equation}
and the dark mode
\begin{equation}
c_{\rm D}
=
\sin\theta\,a
-
e^{i\phi}\cos\theta\,b,
\end{equation}
%
satisfying $[c_{\rm B},c_{\rm D}^\dagger]=0$.
Since the collective jump operator is
$z_{\theta,\phi}=c_{\rm B}$, the engineered reservoir
directly damps the bright mode but does not act on the
orthogonal dark mode. 
The mode $c_{\rm D}$ is exactly dark with respect to
the engineered collective reservoir, since it is
orthogonal to the jump operator $z_{\theta,\phi}=c_{\rm B}$.
Notice that this does not mean that it is lossless or that it remains
an eigenmode of the full dynamics. Local losses give it
a finite lifetime, while the charger-selective gain,
unequal local damping, or detuning can mix the bright
and dark superpositions. We therefore call the
least-damped, or unstable, dynamical eigenmode
\emph{quasi-dark}: it is predominantly composed of
$c_{\rm D}$ but generally retains a small bright-mode
component.

To quantify this distinction, let
$v_+=(v_a,v_b)^T$ denote the normalized right
eigenvector of the full drift matrix associated with
its largest eigenvalue real part. We define its
bright-mode participation as
\begin{equation}
\epsilon_{\rm B}
=
\left|
\cos\theta\,v_a
+
e^{i\phi}\sin\theta\,v_b
\right|^2.
\end{equation}
For an ideal dark eigenmode,
$\epsilon_{\rm B}=0$.
At our reference operating point,
$\epsilon_{\rm B}\simeq 7.3\times10^{-4}$,
showing that the growing mode is almost entirely
dark. Its coupling to the collective reservoir
is therefore strongly suppressed, while the
incoherent gain compensates the remaining losses
and makes this quasi-dark mode unstable.

This distinction is important when interpreting the damping of the battery. The diagonal term of the drift matrix contains the rate $\kappa_b+\Gamma_{\rm col}\sin^2\theta$, which is, however, not the decay rate of the collective dark mode. The common reservoir also generates an off-diagonal dissipative coupling between the charger and battery. In particular, its contribution to the battery amplitude obeys
\begin{equation}
\left.\frac{d\langle b\rangle}{dt}\right|_{\rm col}
=
-\frac{\Gamma_{\rm col}}{2}
e^{-i\phi}\sin\theta
\left(
\cos\theta\,\langle a\rangle
+
e^{i\phi}\sin\theta\,\langle b\rangle
\right).
\end{equation}
For a coherent excitation satisfying
$\langle c_{\rm B}\rangle=0$, the two contributions
cancel exactly. The collective reservoir then produces
no damping of the coherent dark-mode amplitude, even
though the diagonal battery damping coefficient is
large. This cancellation is the physical origin of
the dark-mode protection. 

A vanishing coherent amplitude
$\langle c_{\rm B}\rangle=0$ does not, by itself,
mean that all collective dissipation disappears.
Bright-mode fluctuations can remain, with
$\langle c_{\rm B}^\dagger c_{\rm B}\rangle>0$.
Complete protection from the collective dissipator
requires the absence of bright-mode excitations.
Moreover, local losses, detuning, and the
charger-selective gain can give the full dynamical
eigenmode a finite linewidth or mix the bright and
dark superpositions. We therefore refer to the
selected charging mode as quasi-dark.

Such a distinction can be seen directly after switching
off the coherent drive and incoherent gain. At resonance,
with $\kappa_{a,\downarrow}=\kappa_b=\kappa_0$, the
bright and dark modes become exact eigenmodes of the
remaining dissipative dynamics, with amplitude
eigenvalues
\begin{equation}
\lambda_{\rm D}=-\frac{\kappa_0}{2},
\qquad
\lambda_{\rm B}
=-\frac{\kappa_0+\Gamma_{\rm col}}{2}.
\end{equation}
For $\Gamma_{\rm col}/\kappa_0=17.165$, these give
$\lambda_{\rm D}/\kappa_0=-0.5$ and
$\lambda_{\rm B}/\kappa_0=-9.0825$. Thus, the
large collective decay rapidly suppresses the bright
mode while leaving the dark mode subject only to
local loss. When the incoherent gain is restored,
the dynamical eigenmodes change, and the quasi-dark
branch can become unstable while the bright branch
remains strongly damped.

The reduced instability threshold follows from this
same mechanism. In the strong-collective-dissipation
regime,
\begin{equation}
\Gamma_{\rm col}\sin^2\theta\gg\kappa_b,
\end{equation}
the leading quasi-dark eigenvalue near the instability
threshold is approximately
\begin{equation}
\lambda_{\rm D}
\simeq
-\frac{1}{2}
\left[
(\kappa_{a,\downarrow}-\kappa_{a,\uparrow})
\sin^2\theta
+
\kappa_b\cos^2\theta
\right].
\end{equation}
The gain acts on the charger component of the
quasi-dark mode, while local battery loss acts on
its battery component. The instability begins when
the gain compensates these residual losses. Accordingly,
Eq.~\eqref{eq:kup_crit_diss_normalized} reduces to
\begin{equation}
\kappa_{a,\uparrow}^{\rm crit,diss}
\simeq
\kappa_{a,\downarrow}
+
\kappa_b\cot^2\theta.
\label{eq:kup_critical_dark_limit}
\end{equation}
This explains why dissipative coupling can become unstable at a lower pump rate than coherent coupling.

\begin{figure}[t]
\centering
\includegraphics[width=0.5\columnwidth]{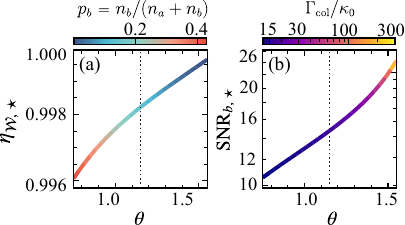}
\caption{Operating-point tradeoff for the dissipative-coupling scheme.
(a) Ergotropy fraction $\eta_{\mathcal W,\star}$,
with curve color indicating the battery occupation
fraction $p_b=n_b/(n_a+n_b)$.
(b) Work-like SNR ${\rm SNR}_{b,\star}=\mathcal W_{b,\star}/\sigma_{E,b,\star}$, with curve color indicating the collective decay rate $\Gamma_{\rm col}/\kappa_0$.
Both quantities are evaluated at the first-passage
target $n_\star=1000$. For each $\theta$, $\Gamma_{\rm col}$ and $\kappa_{a,\uparrow}$ are adjusted to maintain $g/\kappa_0=3.2$ and $\lambda_{\max}/\kappa_0=0.05$. The vertical dotted lines mark the reference operating point $\theta=1.15$. All other parameters are the same as those in Fig.~4 of the main text.\justifying}
\label{fig:theta_tradeoff}
\end{figure}

However, there is a price for lowering the threshold. As $\theta$ approaches $\pi/2$, the dark mode becomes increasingly charger-like, lowering the gain threshold but reducing its battery participation. Consequently, suppressing the instability threshold does not guarantee faster or more efficient battery charging. At fixed coupling strength and growth rate, the charging power decreases as the dark-mode contribution to the battery becomes smaller, as confirmed by our finite-capacity analysis. For an ideal dark-mode excitation, the battery occupation fraction is $p_b^{\rm D}=\cos^2\theta$, which tends to zero in
this limit. At fixed $\Gamma_{\rm col}$, the effective
charger--battery coupling
$|g|=(\Gamma_{\rm col}/2)\sin\theta\cos\theta$ also vanishes. Consequently, the threshold approaches $\kappa_{a,\downarrow}$, but the charger and battery decouple and energy can no longer be transferred to
the battery through the engineered reservoir. Maintaining a finite $|g|$ instead requires an increasingly large $\Gamma_{\rm col}$. Thus, a lower
instability threshold does not automatically imply larger battery occupation, higher charging power, or better charging efficiency. This tradeoff is quantified in Fig.~\ref{fig:theta_tradeoff}.

The stability condition obtained above is closely related to the zero-linewidth condition for bound states in the continuum (BICs) in dissipatively coupled systems. This correspondence follows from identifying $\Gamma_{\rm BIC}=|g|$, $\alpha=\gamma_a$, $\beta=\gamma_b$, and $\Delta_H=2\Delta$, where
$\alpha$, $\beta$, and $\Delta_H$ follow the notation of Ref.~\cite{PhysRevLett.125.147202}. The physical interpretation is nevertheless different. In a passive BIC, destructive interference suppresses radiative decay. In our gain-assisted system, the
collective interference suppresses the decay of the quasi-dark mode, while the incoherent pump compensates its residual losses. The resulting instability threshold marks the onset of amplification.

An exceptional point (EP), by comparison, requires the coalescence of both eigenvalues and eigenvectors. From Eq.~\eqref{eq:splus_general}, the EP condition is
\begin{equation}
(\Delta\gamma+i\Delta)^2+\Pi=0.
\label{EPcondition}
\end{equation}
For dissipative coupling, $\Pi=|g|^2>0$, so an EP requires $\Delta\gamma=0$ and $|\Delta|=|g|$. These conditions characterize a spectral degeneracy, whereas the instability threshold is determined by the vanishing real part of the dominant eigenvalue.


\suppnote{Gaussian equations and charging observables}
\label{sec:gaussian_A_D}

To calculate the charging dynamics and figures of merit discussed in the main text, we employ the Gaussian-state formalism. Since the Hamiltonian is quadratic and the Lindblad operators are linear in bosonic creation and annihilation operators, an initially Gaussian state remains Gaussian throughout the evolution. The dynamics can therefore be fully characterized by its first moments and covariance matrix.

\suppsubsection{First-moment vector and covariance matrix}

We introduce the quadrature vector
\begin{equation}
\bm r=(x_a,p_a,x_b,p_b)^{\mathsf T},
\qquad
x_j=\frac{j+j^\dagger}{\sqrt2},
\qquad
p_j=\frac{j-j^\dagger}{i\sqrt2},
\end{equation}
where $j=a,b$ and $[x_j,p_j]=i$. The corresponding symplectic matrix is \(\mathcal S=\operatorname{diag}(S_2,S_2)\), with \(S_2=\bigl(\begin{smallmatrix}0&1\\-1&0\end{smallmatrix}\bigr)\). 
We define the first-moment vector and covariance matrix as
\begin{equation}
\bm m=\langle\bm r\rangle,
\qquad
V_{ij}=\frac12\langle\Delta r_i\Delta r_j+\Delta r_j\Delta r_i\rangle,
\end{equation}
where $\Delta r_i=r_i - \langle r_i\rangle$. The vacuum state has the covariance matrix \(V_{\rm vac}=\mathbb I/2\)~\cite{Serafini2017QuantumCV}. For the master equation used in the main text, their evolution is governed by the coupled equations
\begin{equation}
\dot{\bm m}=A\bm m+\mathcal S\bm d,
\qquad
\dot V=AV+VA^{\mathsf T}+D.
\label{eq:Lyapunov_general}
\end{equation}
Here
\begin{equation}
A=
\begin{pmatrix}
-\kappa_a^{\rm eff}/2 & \delta_d+\Delta & J_I-\Gamma_{\rm col}u/2 & J_R+\Gamma_{\rm col}v/2\\
-(\delta_d+\Delta) & -\kappa_a^{\rm eff}/2 & -J_R-\Gamma_{\rm col}v/2 & J_I-\Gamma_{\rm col}u/2\\
-J_I-\Gamma_{\rm col}u/2 & J_R-\Gamma_{\rm col}v/2 & -\kappa_b^{\rm eff}/2 & \delta_d-\Delta\\
-J_R+\Gamma_{\rm col}v/2 & -J_I-\Gamma_{\rm col}u/2 & -(\delta_d-\Delta) & -\kappa_b^{\rm eff}/2
\end{pmatrix}
\label{eq:A_explicit}
\end{equation}
is the quadrature drift matrix, with
\begin{equation}
\begin{aligned}
\kappa_a^{\rm eff}
&=
\kappa_{a,\downarrow}-\kappa_{a,\uparrow}+\Gamma_{\rm col}\cos^2\theta,
\qquad
\kappa_b^{\rm eff}
=
\kappa_b+\Gamma_{\rm col}\sin^2\theta,
\\
u
&=
\sin\theta\cos\theta\cos\phi,
\qquad
v
=
\sin\theta\cos\theta\sin\phi,
\end{aligned}
\end{equation}
and \(J=J_R+iJ_I\). This expression includes both coupling mechanisms considered in the main text: the coherent-coupling scheme is obtained by setting \(\Gamma_{\rm col}=0\), whereas the dissipative-coupling scheme corresponds to \(J=0\).
\begin{equation}
D=
\begin{pmatrix}
D_a&0&D_u&-D_v\\
0&D_a&D_v&D_u\\
D_u&D_v&D_b&0\\
-D_v&D_u&0&D_b
\end{pmatrix},
\label{eq:D_explicit}
\end{equation}
is the diffusion matrix that describes fluctuations introduced by the dissipation channels, with 
\begin{equation}
\begin{aligned}
D_a
&=
\frac{\kappa_{a,\downarrow}+\kappa_{a,\uparrow}+\Gamma_{\rm col}\cos^2\theta}{2},
\qquad
D_b
=
\frac{\kappa_b+\Gamma_{\rm col}\sin^2\theta}{2},
\\
D_u
&=
\frac{\Gamma_{\rm col}}{2}\sin\theta\cos\theta\cos\phi,
\qquad
D_v
=
\frac{\Gamma_{\rm col}}{2}\sin\theta\cos\theta\sin\phi.
\end{aligned}
\end{equation}
The diagonal entries of \(D\) describe fluctuations associated with the local gain and loss channels as well as the collective reservoir. In contrast, its off-diagonal blocks describe correlated quantum noise arising from the common reservoir. Importantly, the same collective dissipative channel generates both the off-diagonal damping terms in \(A\) and the noise correlations in \(D\).
The coherent drive enters through
\begin{equation}
\bm{d}=
\bigl(\sqrt2\operatorname{Re}[\mathcal E],\sqrt2\operatorname{Im}[\mathcal E],0,0\bigr)^{\mathsf T}.
\end{equation}


Together, these equations determine the first moments and covariance matrix needed to evaluate the battery energy, ergotropy, charging efficiency, and work-like SNR, as discussed in the following subsection. For the stability scans, we fix $\delta_d=0$ and vary $\Delta$. For the charging dynamics, we set $\omega_a=\omega_b$ ($\Delta=0$); optimizing the coherent driving frequency then changes only $\delta_d$.

%




\suppsubsection{Figures of merit and finite-capacity performance}

To quantify the charging performance discussed in the main text, we first introduce the battery-energy and work-quality measures, followed by the corresponding charging powers and active-energy costs.

The energy stored in the battery is defined with respect to its bare Hamiltonian $H_b=\omega_b b^\dagger b$, as
\begin{equation}
E_b(t)=\omega_b\langle b^\dagger b\rangle.
\end{equation}
The maximum work extractable from the battery is quantified by its ergotropy~\cite{Campaioli2024Colloquium} 
\begin{equation}
    \mathcal{W}_{b} \equiv \mathrm{Tr}[H_b\rho_b] - \mathrm{Tr}[H_b\rho_b^{\rm pas}],
    \label{Wb_exact}
\end{equation}
where $\rho_b^{\rm pas}$ is the passive state corresponding to the reduced battery state $\rho_b$, obtained by assigning the eigenvalues of $\rho_b$ (in descending order) to the eigenstates of $H_b$ (in ascending order). 
For a general single-mode Gaussian state characterized by a local covariance matrix $V_b$ and first-moment vector $\bm m_b=(\langle x_b\rangle,\langle p_b\rangle)^{\mathsf T}$, the ergotropy takes the exact form
\begin{equation}
\mathcal W_b
=
\frac{\omega_b}{2}
\left(
\operatorname{Tr}V_b
+
\bm m_b^{\mathsf T}\bm m_b
-
2\sqrt{\det V_b}
\right).
\label{eq:gaussian_battery_ergotropy}
\end{equation}
This expression identifies two contributions to the ergotropy. The term $\omega_b\bm m_b^{\mathsf T}\bm m_b/2$ represents the work stored in the coherent displacement, while $\omega_b(\operatorname{Tr}V_b-2\sqrt{\det V_b})/2$ represents the work associated with squeezing. 
For the number-conserving Hamiltonian, phase-insensitive Lindblad channels, and vacuum or thermal initial states considered here, the dynamics generates no squeezing correlations. The reduced battery therefore remains a displaced thermal state, with
\begin{equation}
V_b(t)
=
\left(n_b^{\rm inc}(t)+\frac12\right)\mathbb I_2,
\end{equation}
where $n_b^{\rm inc}(t)=\langle b^\dagger(t) b(t)\rangle-|\langle b(t)\rangle|^2$ denotes the occupation of fluctuations around the coherent displacement. Consequently, 
\begin{equation}
\mathcal W_b(t)
=
\omega_b|\langle b(t)\rangle|^2,
\qquad
E_b^{\rm pas}(t)
=
\omega_b n_b^{\rm inc}(t).
\label{eq:ergotropy_equals_displacement_energy}
\end{equation}
Thus, the coherent displacement carries the extractable work, while the incoherent occupation contributes to the passive energy.


To evaluate the work-like SNR introduced in the main text, we calculate the energy variance for the displaced thermal battery state. Writing $b=\beta+d$, where $\beta=\langle b\rangle$, $\langle d\rangle=\langle dd\rangle=0$, and $\langle d^\dagger d\rangle=n_b^{\rm inc}$, the photon-number operator becomes
\begin{equation}
n_b=b^\dagger b=|\beta|^2+\beta^*d+\beta d^\dagger+d^\dagger d.
\label{eq:sm_displaced_number}
\end{equation}
For thermal fluctuations, $\operatorname{Var}(d^\dagger d)=n_b^{\rm inc}(n_b^{\rm inc}+1)$, and the cross terms between the linear and quadratic fluctuation terms vanish. We therefore obtain
\begin{equation}
\begin{aligned}
\operatorname{Var}(n_b)
&=\langle n_b^2\rangle-\langle n_b\rangle^2\\
&=n_b^{\rm inc}(n_b^{\rm inc}+1)+(2n_b^{\rm inc}+1)|\beta|^2,\\
\sigma_{E,b}&=\omega_b\sqrt{n_b^{\rm inc}(n_b^{\rm inc}+1)+(2n_b^{\rm inc}+1)|\beta|^2}.
\end{aligned}
\label{eq:sm_energy_variance_displaced}
\end{equation}
The ergotropy fraction and work-like SNR are then given by
\begin{equation}
\begin{aligned}
\frac{\mathcal{W}_b(t)}{E_b(t)}&=\frac{|\beta(t)|^2}{|\beta(t)|^2+n_b^{\rm inc}(t)},\\
\mathrm{SNR}_b(t)&=\frac{|\beta(t)|^2}
{\sqrt{n_b^{\rm inc}(t)[n_b^{\rm inc}(t)+1]+[2n_b^{\rm inc}(t)+1]|\beta(t)|^2}}.
\end{aligned}
\label{eq:SNR_def}
\end{equation}
For a given coherent displacement, increasing the incoherent occupation enhances the battery-energy fluctuations and reduces the work-like SNR.

For the finite-capacity analysis in the main text, we consider a prescribed target occupation $n_\star$. The \emph{first-passage charging time}, i.e., 
\begin{equation}
t_{\rm ch}(n_\star)
=
\inf\{t\ge0:\langle n_b(t)\rangle\ge n_\star\},
\label{eq:finite_capacity_charging_time}
\end{equation}
is the earliest instant at which the total battery occupation reaches this target. At this stopping time, we quantify the charging schemes with the work-like 
\(
\mathrm{SNR}_{b,\star}
=
\mathrm{SNR}_b[t_{\rm ch}(n_\star)]
\). 
We also define the average ergotropy power as
\begin{equation}
P_{\mathcal W,\star}
=
\frac{\mathcal W_b[t_{\rm ch}(n_\star)]}
{t_{\rm ch}(n_\star)},
\label{eq:finite_average_power}
\end{equation}
which is the extractable work gained during the charging process divided by the corresponding charging time.

We now turn to the \emph{active-energy cost} of charging, i.e., the energies supplied through the active channels. In general, the contribution of a dynamical channel $\mathcal L_\alpha$ to the rate of change of a system energy associated with a reference Hamiltonian $H_{\rm ref}$ is defined as
\begin{equation}
P_\alpha(t)
\equiv
\left.
\frac{d\langle H_{\rm ref}\rangle}{dt}
\right|_\alpha
=
\operatorname{Tr}
\left[
H_{\rm ref}\mathcal L_\alpha[\rho(t)]
\right].
\label{eq:channel_power_definition}
\end{equation}

For the active channels of our model, the relevant terms in the master-equation generator are
\begin{equation}
\begin{aligned}
\mathcal L_{\rm drive}[\rho]
&=-i[H_{\rm drive},\rho],
\\
\mathcal L_{\rm gain}[\rho]
&=\kappa_{a,\uparrow}\mathcal D[a^\dagger]\rho,
\end{aligned}
\end{equation}
where $H_{\rm drive}=\mathcal E a^\dagger+
\mathcal E^*a$. These are instantaneous generators acting on the
same state $\rho(t)$, which evolves under the full master equation. Their contributions to the bare charger-energy change can be evaluated separately using
Eq.~\eqref{eq:channel_power_definition}, with
$H_{\rm ref}=H_a=\omega_a a^\dagger a$.
The resulting powers quantify the energy transferred
to the bare charger and do not represent the entire energy cost of implementing these
channels. In particular, for an off-resonant coherent drive, the laboratory-frame source-work rate involves $\omega_d$ instead of $\omega_a$.
The microscopic cost of maintaining the gain reservoir
is also excluded. Within the bare-energy convention,
the two channel contributions are
\begin{equation}
\begin{aligned}
P_{\rm drive}(t)
&=
-2\omega_a\operatorname{Im}
\left[\mathcal E^*\langle a(t)\rangle\right],
\\
P_{\rm gain}(t)
&=
\omega_a\kappa_{a,\uparrow}
\bigl(\langle n_a(t)\rangle+1\bigr),
\\
E_{\rm act}(t)
&=E_{\rm gain}(t) + E_{\rm drive}(t) =
\int_0^t\!dt'\,
\left(P_{\rm drive}(t')+P_{\rm gain}(t')\right),
\end{aligned}
\label{eq:net_active_energy_definitions}
\end{equation}
with which one can define the finite-capacity \emph{charging efficiency}
\begin{equation}
\eta_{{\rm ch},\star}
=
\frac{\mathcal W_b[t_{\rm ch}(n_\star)]}
{E_{\rm act}[t_{\rm ch}(n_\star)]}.
\label{eq:finite_efficiency}
\end{equation}

To clarify these expressions, we first consider the incoherent gain channel. Its contribution to the charger occupation follows from the bosonic commutation relation $[a,a^\dag]=1$:
\begin{equation}
\left.
\frac{d\langle n_a\rangle}{dt}
\right|_{\rm gain}
=
\kappa_{a,\uparrow}
\bigl(\langle n_a\rangle+1\bigr).
\label{eq:gain_occupation_rate}
\end{equation}
The term proportional to $\langle n_a\rangle$ describes the stimulated amplification, while the additional $1$ represents the spontaneous excitation injection. Thus, the gain reservoir can supply energy even when the charger is initially empty, and the injection rate increases with its occupation. Multiplying Eq.~\eqref{eq:gain_occupation_rate} by $\omega_a$ gives $P_{\rm gain}$, the instantaneous energy delivered to the charger by the incoherent pump.

The coherent-drive power has a different character. It depends on the phase of the charger response relative to the applied field and can therefore have either sign. Positive $P_{\rm drive}$ describes energy transferred from the coherent drive into the charger, while negative $P_{\rm drive}$ describes the reverse transfer. As shown in the main text, our signed-input convention credits this backflow when evaluating $E_{\rm act}$. 

Importantly, neither $P_{\rm drive}$ nor $P_{\rm gain}$ should be identified with the battery charging power $P_b$. The battery energy also changes through the charger--battery coupling and dissipative processes, so the energy delivered to the charger need not be stored in the battery. Moreover, the active-energy cost measures the energy transferred through the specified active channels, not the full physical cost of operating their sources. In particular, it does not include the microscopic cost of maintaining population inversion or engineering the collective reservoir, nor does it separately account for the interaction energy. The resulting charging efficiency is therefore a model-dependent energy-cost measure rather than a complete thermodynamic efficiency. The gross-input convention discussed in the End Matter instead counts the total coherent-field energy delivered to the device, without crediting any energy returned to the source.

Finally, to ensure a consistent comparison between the coherent- and dissipative-coupling schemes, we use the following matched conditions:
\begin{equation}
\frac{J}{\kappa_0}
=
\frac{g}{\kappa_0}
=3.2,
\qquad
\frac{\lambda_{\max}}{\kappa_0}=0.05,
\qquad
\frac{\kappa_{a,\downarrow}}{\kappa_0}
=
\frac{\kappa_b}{\kappa_0}=1,
\qquad
\frac{\mathcal E}{\kappa_0}=10.
\label{eq:matched_growth_benchmark}
\end{equation}
Here \(\lambda_{\max}=\max_j[\operatorname{Re}{\lambda_j}]\) denotes the largest drift-eigenvalue real part.
The coherent-coupling scheme has $\kappa_{a,\uparrow}/\kappa_0=2.2$ and $J/\kappa_0=3.2$. The dissipative-coupling scheme has $\theta=1.15$, $\Gamma_{\rm col}/\kappa_0=17.165$, $\kappa_{a,\uparrow}/\kappa_0=1.30457$, and $g/\kappa_0=3.2$. For the dissipative-coupling scheme, we always set $\phi=0$ without loss of generality. These parameters yield matched off-diagonal coupling magnitudes ($|J|=|g|$) and the same maximal growth rate $\lambda_{\max}$. All simulations start from the vacuum unless stated otherwise.

%

\suppnote{Operating-point tradeoff}\label{sec:OP_tradeoff}

In the main text, we use $\theta=1.15$ as the operating point of the dissipative-coupling scheme. We now examine how the charging quality depends on the composition of the dissipative coupling. In Fig.~\ref{fig:theta_tradeoff}, we vary $\theta$ and adjust $\Gamma_{\rm col}$ and $\kappa_{a,\uparrow}$ such that the dissipative coupling strength $g/\kappa_0=3.2$ and the maximal growth rate $\lambda_{\max}/\kappa_0=0.05$ remain fixed. This isolates the effect of rotating the collective mode from changes in coupling strength or amplification rate. 

It shows that both the ergotropy fraction and the work-like SNR improve as $\theta$ approaches $\pi/2$. However, this improvement is accompanied by a decreasing battery participation $p_b=n_b/(n_a+n_b)$, as indicated by the color scale, which implies that the unstable collective mode becomes progressively more dominated by the charger. Moreover, as $\theta$ increases toward $\pi/2$, maintaining a fixed $|g|$ requires a rapidly growing collective decay rate $\Gamma_{\rm col}$. Therefore, the optimal charging quality is only obtained at the cost of weaker energy localization in the battery and stronger reservoir engineering. The operating point $\theta=1.15$ used in Figs.~3 and 4 of the main text (indicated by the vertical dotted lines in Fig.~\ref{fig:theta_tradeoff}) provides a representative compromise between these competing requirements. 


\suppnote{Finite-gain implementation}
\label{sec:saturable_gain}

The linear gain channel \(\kappa_{a,\uparrow}\mathcal D[a^\dagger]\) used in the main text describes amplification without accounting for depletion of the gain medium. As a result, it predicts unbounded exponential growth in the unstable regime. To examine how gain saturation limits this growth, we replace the phenomenological gain channel with an incoherently pumped two-level emitter coupled to the charger. For the dissipative-coupling scheme \(J=0\), the resulting master equation reads
\begin{equation}
\begin{split}
\dot\rho_{\rm sat}
={}&
-i[H_{\rm sat},\rho_{\rm sat}]
+\kappa_{a,\downarrow}\mathcal D[a]\rho_{\rm sat}
+\kappa_b\mathcal D[b]\rho_{\rm sat}
+\Gamma_{\rm col}\mathcal D[z_{\theta,\phi}]\rho_{\rm sat}
\\
&
+P_\sigma\mathcal D[\sigma_+]\rho_{\rm sat}
+\gamma_\sigma\mathcal D[\sigma_-]\rho_{\rm sat}
+\frac{\gamma_{\sigma,\varphi}}{2}\mathcal D[\sigma_z]\rho_{\rm sat},
\end{split}
\label{eq:saturable_master_equation}
\end{equation}
where the Hamiltonian is
\begin{equation}
H_{\rm sat}
={}
(\delta_d+\Delta)a^\dagger a
+(\delta_d-\Delta)b^\dagger b
+\Delta_\sigma\sigma_+\sigma_-
+g_\sigma^{(1)}(a^\dagger\sigma_-+a\sigma_+)
+\mathcal E a^\dagger+\mathcal E^*a.
\label{eq:saturable_hamiltonian}
\end{equation}
Here \(\sigma_\pm\) and \(\sigma_z\) are the raising/lowering and inversion operators of the emitter, respectively. The parameters $P_{\sigma}$, $\gamma_{\sigma}$, and $\gamma_{\sigma,\varphi}$ describe its incoherent pumping, relaxation, and pure dephasing, while $\Delta_{\sigma}$ and $g_{\sigma}^{(1)}$ denote the emitter's detuning and its coupling strength to the charger mode. Unlike the linear model, the gain is now supplied by the pumped emitter, whose population inversion can be depleted as the charger is amplified.

To describe this saturation mechanism at the mean-field level, we introduce \(\alpha=\langle a\rangle\), \(\beta=\langle b\rangle\), \(s=\langle\sigma_-\rangle\), and \(w=\langle\sigma_z\rangle\). Factorizing the correlations between the bosonic modes and the emitter gives the Maxwell--Bloch equations~\cite{Cartar2017}
\begin{equation}
\begin{aligned}
\dot\alpha
&=
-\left[i(\delta_d+\Delta)+\frac{\kappa_{a,\downarrow}+\Gamma_{\rm col}\cos^2\theta}{2}\right]\alpha
-\frac{\Gamma_{\rm col}}{2}e^{i\phi}\sin\theta\cos\theta\,\beta
-i g_\sigma^{(1)} s-i\mathcal E,
\\
\dot\beta
&=
-\left[i(\delta_d-\Delta)+\frac{\kappa_b+\Gamma_{\rm col}\sin^2\theta}{2}\right]\beta
-\frac{\Gamma_{\rm col}}{2}e^{-i\phi}\sin\theta\cos\theta\,\alpha,
\\
\dot s
&=
-(\gamma_\perp+i\Delta_\sigma)s+i g_\sigma^{(1)}\alpha w,
\\
\dot w
&=
-\gamma_\parallel(w-w_0)
+2i g_\sigma^{(1)}(\alpha^*s-\alpha s^*).
\end{aligned}
\label{eq:saturable_maxwell_bloch}
\end{equation}
Here $\gamma_\parallel$ and $\gamma_\perp$ are the longitudinal and transverse relaxation rates of the emitter, respectively, and $w_0$ is its steady-state inversion in the absence of coupling to the charger:
\begin{equation}
\gamma_\parallel=P_\sigma+\gamma_\sigma,
\qquad
\gamma_\perp=\frac{\gamma_\parallel}{2}+\gamma_{\sigma,\varphi},
\qquad
w_0=\frac{P_\sigma-\gamma_\sigma}{P_\sigma+\gamma_\sigma}.
\label{eq:saturable_tls_rates}
\end{equation}
Population inversion requires \(P_\sigma>\gamma_\sigma\). In the weak-field regime, the emitter inversion remains close to \(w_0\), allowing it to provide effective gain to the charger. As the charger field grows, stimulated emission depletes the inversion and reduces the available gain, eventually limiting further amplification.

For a consistent comparison with the phenomenological linear-gain model, we choose the emitter-charger coupling \(g_\sigma^{(1)}\) such that the two models have the same small-signal instability growth rate. Specifically, we linearize the homogeneous Maxwell--Bloch equations around the unsaturated inversion \(w=w_0\) and determine \(g_\sigma^{(1)}\) by requiring that the largest real part of the resulting eigenvalues equals the prescribed growth rate \(\lambda_\star\) of the linear-gain benchmark.

The saturation mechanism can be understood analytically in the fast-emitter regime. By adiabatically eliminating the emitter polarization and assuming that the inversion follows its quasi-steady state, we obtain
\begin{equation}
 w
 \simeq
 \frac{w_0}{1+|\alpha|^2/n_{\rm sat}},
 \qquad
 n_{\rm sat}
 = \frac{\gamma_\parallel(\gamma_\perp^2+\Delta_\sigma^2)}
 {4\left(g_\sigma^{(1)}\right)^2\gamma_\perp}.
\label{eq:saturation_occupation}
\end{equation}
Here $n_{\rm sat}$ is the characteristic coherent charger occupation at which the emitter inversion is reduced to half its unsaturated value $w_0$. Substituting this expression into the equation for the charger amplitude yields the occupation-dependent gain rate
\begin{equation}
\kappa_{a,\uparrow}^{\rm sat}(n)
=
\frac{\kappa_{a,\uparrow}^{(0)}}{1+|\alpha|^2/n_{\rm sat}},
\qquad
\kappa_{a,\uparrow}^{(0)}
=
\frac{2\left(g_\sigma^{(1)}\right)^2\gamma_\perp w_0}{\gamma_\perp^2+\Delta_\sigma^2},
\label{eq:saturable_gain_rate}
\end{equation}
where $\kappa_{a,\uparrow}^{(0)}$ denotes the unsaturated gain rate. Thus, the effective gain approaches its small-signal value for $|\alpha|^2 \ll n_{\rm sat}$ and decreases continuously as the charger field grows. This provides a simple explanation for how gain depletion suppresses the unbounded amplification predicted by the linear model.

To verify this mechanism beyond the mean-field approximation, we numerically solve the full master equation for a single pumped emitter [cf. Eq.~\eqref{eq:saturable_master_equation}]. The parameters are chosen as 
\begin{equation}\label{SM_parameter1}
\frac{\kappa_{a,\downarrow}}{\kappa_0}
=
\frac{\kappa_b}{\kappa_0}=5,
\qquad
\frac{\Gamma_{\rm col}}{\kappa_0}=200,
\qquad
\theta=\arctan 2,
\qquad
\frac{\mathcal E}{\kappa_0}=2,
\end{equation}
and
\begin{equation}\label{parametersS46}
\frac{P_\sigma}{\kappa_0}=16,\qquad
\frac{\gamma_\sigma}{\kappa_0}=3,\qquad
\frac{g_\sigma^{(1)}}{\kappa_0}=6.89,\qquad
\Delta_\sigma=\Delta=\delta_d=\gamma_{\sigma,\varphi}=0.
\end{equation}
The emitter--charger coupling $g_\sigma^{(1)}$ is
chosen such that the maximal growth rate of the pumped-emitter model matches the prescribed value \(\lambda_\star/\kappa_0=0.2\) of the linear-gain benchmark. 
%
\begin{figure}[t]
\centering
\includegraphics[width=0.70\textwidth]{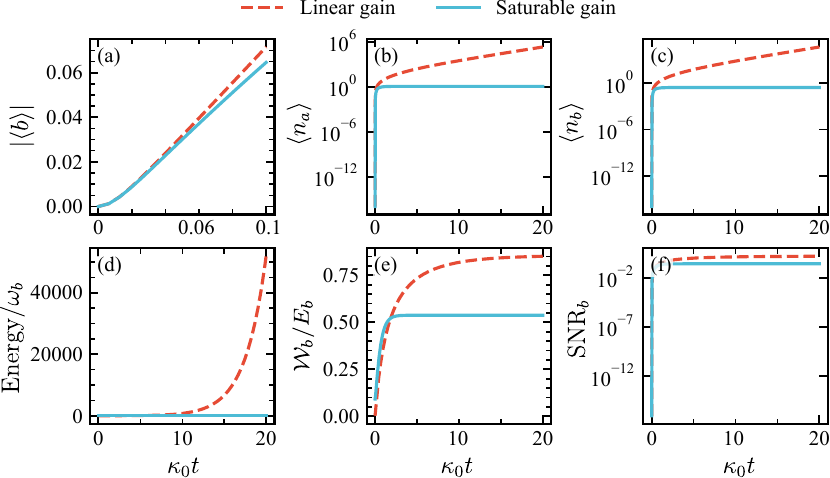}
\caption{\justifying Single-emitter gain saturation and charging dynamics.
Comparison between the phenomenological linear-gain model
(red dashed) and the pumped-emitter model (cyan solid),
with the same small-signal growth rate
$\lambda_\star/\kappa_0=0.2$.
(a) Early-time coherent battery amplitude $|\langle b\rangle|$.
(b), (c) Mean occupations $\langle n_a\rangle$ and
$\langle n_b\rangle$, respectively.
(d) Stored battery energy $E_b/\omega_b$.
(e) Ergotropy fraction $\mathcal{W}_b/E_b$.
(f) Work-like SNR ${\rm SNR}_b=\mathcal W_b/\sigma_{E,b}$.
The pumped-emitter model initially follows the linear-gain
dynamics but eventually saturates due to the emitter-inversion
depletion, whereas the linear model exhibits unbounded growth. All other parameters are specified in the text [Eqs.~\eqref{SM_parameter1} and \eqref{parametersS46}]. }
\label{fig:saturable_dynamics}
\end{figure}
%


Figure~\ref{fig:saturable_dynamics} compares the resulting charging dynamics with those predicted by the linear-gain model at the same small-signal growth rate. Both models exhibit similar exponential growth at early times, but the pumped-emitter model eventually saturates at a finite steady-state battery occupation, while the linear-gain model continues to grow without bound. Although the gain saturation mechanism reduces both the ergotropy fraction and work-like SNR relative to the linear-gain model, increasing the number of pumped emitters may improve these quantities by supporting a larger steady-state coherent battery occupation, as shown in Fig.~6 of the main text.
For the single-emitter master equation~\eqref{eq:saturable_master_equation}, the reduced battery state may become non-Gaussian. Accordingly, the ergotropy is directly obtained from Eq.~\eqref{Wb_exact}, with \(\rho_b=\operatorname{Tr}_{a,\sigma}[\rho_{\rm sat}]\), while the battery-energy standard deviation is obtained from $\sigma_{E,b}=\omega_b\sqrt{\operatorname{Tr}(\rho_b n_b^2)-[\operatorname{Tr}(\rho_b n_b)]^2}$.

\begin{figure}[t]
\centering
\includegraphics[width=0.70\textwidth]{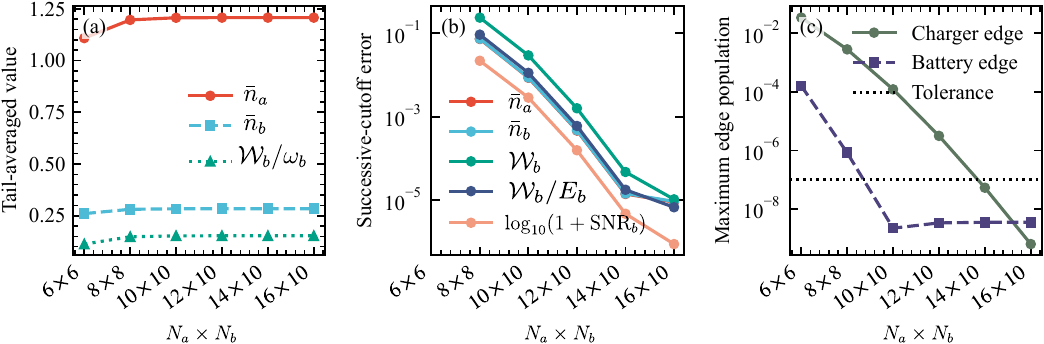}
\caption{Fock-space truncation convergence of the exact single-emitter master-equation calculation.
(a) Tail-averaged charger and battery occupations
$\bar n_a$ and $\bar n_b$, and battery ergotropy $\overline{\mathcal W}_b/\omega_b$.
(b) Relative changes between successive truncations for $\bar n_a$, $\bar n_b$, $\mathcal{W}_b$, $\mathcal{W}_b/E_b$, and $\log_{10}(1+\mathrm{SNR}_b)$.
(c) Maximum populations at the charger and battery Fock-space boundaries, with the horizontal dotted line indicating the numerical tolerance.
The truncation sequence is $(N_a,N_b)=(6,6),(8,8),(10,10),(12,10),(14,10),(16,10)$.
The results converge with increasing truncation dimensions, supporting the physical origin of the observed gain saturation. All other parameters are the same as those in Fig.~\ref{fig:saturable_dynamics}. \justifying}
\label{fig:saturable_convergence}
\end{figure}

To confirm that this saturation is a physical effect rather than an artifact of the finite Fock-space truncation, we perform the numerical convergence analysis shown in Fig.~\ref{fig:saturable_convergence}. Panel (a) shows that the tail-averaged charger and battery occupations and battery ergotropy converge as the truncation dimensions increase. Panel (b) confirms the convergence of more relevant quantities, including the ergotropy fraction and work-like SNR. Panel (c) further shows that the populations near the Fock-space boundaries become negligible. For the final increment, the relative differences over the entire simulated time interval are below \(1.1\times10^{-5}\) for \(\langle n_a\rangle\), \(\langle n_b\rangle\), and \(\mathcal W_b\), while the populations in the highest two retained Fock levels remain below \(4\times10^{-9}\).

We next extend the saturable-gain model to an ensemble of \(N_\sigma\) identical pumped emitters to investigate how the emitter number affects the attainable battery occupation. In this case, the interaction Hamiltonian is
\begin{equation}
H_{\rm int}^{(N_\sigma)} = g_\sigma^{(N_\sigma)}\sum_{j=1}^{N_\sigma}\left(a^\dagger\sigma_-^{(j)}+a\sigma_+^{(j)}\right),
\end{equation}
with each emitter subject to local relaxation and dephasing. We choose
\begin{equation}
g_\sigma^{(N_\sigma)} = \frac{g_\sigma^{(1)}}{\sqrt{N_\sigma}},
\end{equation}
which keeps the collective emitter–charger coupling (and hence the small-signal gain) fixed as $N_\sigma$ increases. The corresponding emitter-number scaling results are presented in the End Matter. Together, the single-emitter and ensemble calculations demonstrate how gain saturation limits the exponential growth, while larger emitter ensembles enable substantially higher steady-state coherent battery occupations.

\suppnote{Mode selection and phase-locked charging}
\label{sec:snr_argument}

To understand the beating and phase-locked charging dynamics observed in the main text, we examine how the coherent drive and incoherent gain contribute to the amplified modes. As shown in Note~\ref{sec:gaussian_A_D}, the Gaussian equations separate the evolution of the first moments and covariance matrix.
Specifically, their formal solutions are
\begin{equation}
\begin{aligned}
\bm m(t) &= e^{At}\bm m(0) + \int_0^t e^{A(t-s)}\mathcal S\bm d\,ds,
\\
V(t) &= e^{At}V(0)e^{A^{\mathsf T}t} + \int_0^t e^{A(t-s)}D e^{A^{\mathsf T}(t-s)}\,ds.
\end{aligned}
\label{eq:mean_covariance_solutions}
\end{equation}
These expressions highlight the different roles of coherent driving and incoherent gain. The incoherent gain affects both the drift and diffusion matrices but cannot generate a coherent displacement by itself. Thus, \(\bm m(0)=0\) and \(\mathcal E=0\) imply \(\bm m(t)=0\), even in the unstable regime, while the gain-induced fluctuations can still be amplified. By contrast, the coherent drive provides a phase reference and generates a nonzero displacement, which can subsequently be amplified by the unstable dynamics. Table~\ref{tab:sm_pumping_roles} summarizes the charging response for different combinations of coherent driving and incoherent pumping.


%
\begin{center}
\begin{minipage}{\columnwidth}
\centering
\captionof{table}{Roles of coherent driving (CP) and incoherent pumping (IP) for initially vacuum modes. The coupling scheme determines whether an amplified response exhibits hybrid-mode beating or phase locking.}
\label{tab:sm_pumping_roles}
\begin{tabular}{c@{\hspace{1.2em}}c@{\hspace{1.2em}}l}
\toprule
CP & IP & Charging response from vacuum\\
\midrule
Off & Off & \parbox[t]{0.72\columnwidth}{No excitation or coherent displacement.}\\
On & Off & \parbox[t]{0.72\columnwidth}{Phase-defined displacement, without gain-induced exponential growth.}\\
Off & On & \parbox[t]{0.72\columnwidth}{Incoherent occupation can grow, but the first moments remain zero and the battery is passive in this model.}\\
On & On & \parbox[t]{0.72\columnwidth}{The phase-defined displacement can be amplified exponentially when the drift is unstable and the growing mode is excited.}\\
\bottomrule
\end{tabular}
\end{minipage}
\end{center}

The key difference between the two coupling schemes lies in the off-diagonal structure of their drift matrices. As discussed in Note~\ref{app:exceptional_points}, the products of the off-diagonal matrix elements are invariant under local rephasings of \(a\) and \(b\),
\begin{equation}
(H_r^{\rm coh})_{ab}(H_r^{\rm coh})_{ba} = -|J|^2,
\qquad\qquad
(H_r^{\rm diss})_{ab}(H_r^{\rm diss})_{ba} = |g|^2.
\label{eq:gauge_invariant_coupling_products}
\end{equation}
The opposite signs of these products give rise to frequency and linewidth splitting, respectively. To illustrate this difference, we consider the bare-mode-resonant case \(\Delta=0\) with equal effective damping rates \(\gamma_a=\gamma_b=\gamma\). The corresponding eigenvalues then reduce to
\begin{equation}
\lambda_\pm^{\rm coh}=-\gamma\pm i|J|,
\qquad\qquad
\lambda_\pm^{\rm diss}=-\gamma\pm|g|.
\end{equation}
Coherent coupling therefore produces frequency splitting, whereas dissipative coupling produces linewidth splitting between the collective modes. In the coherent-coupling scheme, both unstable modes grow at the same rate and interfere, producing persistent beating in the battery response. In the dissipative-coupling scheme, the common reservoir strongly damps the bright collective mode while leaving a weakly damped quasi-dark mode. At the matched operating point, only the quasi-dark mode is unstable and dominates the dynamics after a short transient. Consequently, the ratio \(\langle b\rangle/\langle a\rangle\) approaches a constant, giving rise to phase-locked amplification.


Although this mode-selection mechanism is gauge invariant, the numerical value of the relative phase $\Delta\varphi$ depends on the mode-phase convention. Write $J=|J|e^{i\chi}$. In the pure coherent-coupling scheme, the transformation $b\mapsto e^{i\chi}b$ makes $J$ real. This does not affect the spectrum and beating dynamics, but shifts the relative phase of the mean amplitudes by $-\chi$. The relative phases in the original and transformed bases are related by \(\Delta\varphi=\Delta\varphi'-\chi\). In the pure dissipative-coupling scheme, the transformation $b\mapsto e^{i\phi}b$ removes the reservoir phase, giving $\Delta\varphi=\pi-\phi$ in the original basis. Changing $\arg\mathcal E$ rotates both mean amplitudes by the same phase and leaves their relative phase unchanged. When both coherent and dissipative couplings are present, their relative phase \(\chi-\phi\) becomes physically relevant and can give rise to nonreciprocal charging, as studied in Ref.~\cite{PhysRevLett.132.210402}.

The phase locking discussed here concerns the coherent first moments, while the first-order intermode coherence also contains contributions from the quantum fluctuations. Writing \(\alpha=\langle a\rangle\) and \(\beta=\langle b\rangle\), we have
\[
\begin{aligned}
\langle a^\dagger b\rangle
&=\alpha^*\beta+
\langle(a-\alpha)^\dagger(b-\beta)\rangle,\\[6pt]
g_{ab}^{(1)}
&=
\frac{\langle a^\dagger b\rangle}
{\sqrt{\langle a^\dagger a\rangle\langle b^\dagger b\rangle}}.
\end{aligned}
\]

The Gaussian framework also allows us to characterize the mixedness of the amplified battery state through its purity. For vacuum or thermal initial states, the reduced battery remains a displaced thermal state,
\[
\rho_b=D(\beta)\rho_{\rm th}(n_b^{\rm inc})D^\dagger(\beta).
\]
Its purity follows from the unitary invariance of \(\operatorname{Tr}(\rho_b^2)\):
\[
\mathcal P_b\equiv\operatorname{Tr}(\rho_b^2)=\frac{1}{2n_b^{\rm inc}+1}.
\]
At fixed time and for a given drift matrix and diffusion matrix, changing the amplitude or phase of the coherent drive changes the displacement but leaves $n_b^{\rm inc}$ (and hence the purity) unchanged. The ergotropy fraction, however, also depends on the coherent displacement. Therefore, a battery state can exhibit a high ergotropy fraction while remaining strongly mixed due to gain-induced fluctuations.

The different mode dynamics thus determines how the coherent displacement develops, while the fluctuation dynamics governs the accompanying passive energy and mixedness. The following note examines how these contributions accumulate over time and determine the work-like SNR.

\suppnote{Temporal source--noise matching and finite-time work quality}\label{Temporal}

\suppsubsection{Response of a general Gaussian network}

To derive the exact finite-time source--noise relation in the main text, we first consider a general network of $N$ bosonic modes with a number-conserving quadratic Hamiltonian, linear phase-insensitive gain and loss, and coherent driving. This framework allows us to distinguish the coherent response generated by the external drive from the fluctuations introduced by the incoherent gain. 

Let $\bm c=(c_1,\ldots,c_N)^{\mathsf T}$, $\bm\alpha=\langle\bm c\rangle$, and define the normally ordered covariance matrix as 
\[
C_{ij}=\left\langle(c_j-\alpha_j)^\dagger(c_i-\alpha_i)\right\rangle.
\] 
For a time-independent complex drift matrix $M$, a coherent source $\bm f$, and a normally ordered noise-injection matrix $K\succeq0$, the equations of motion for the first moments and covariance matrix are
\begin{equation}
\dot{\bm\alpha}=M\bm\alpha+\bm f,\qquad
\dot C=MC+CM^\dagger+K.
\label{eq:sm_source_noise_evolution}
\end{equation}
Unlike the symmetrically ordered quadrature-diffusion matrix \(D\) introduced in Supplementary Note~\ref{sec:gaussian_A_D}, the matrix \(K\) describes the injection of normally ordered fluctuations. In particular, the vacuum loss channels do not contribute directly to $K$, since they do not inject excitations into the system.

For vacuum initial conditions, $\bm\alpha(0)=0$ and $C(0)=0$, the solutions are
\begin{equation}
\begin{aligned}
\bm\alpha(t)&=\int_0^t e^{Mu}\bm f\,du,\\
C(t)&=\int_0^t e^{Mu}Ke^{M^\dagger u}\,du.
\end{aligned}
\label{eq:sm_source_noise_integrals}
\end{equation}
For the \emph{phase-insensitive} dynamics considered here, the reduced battery remains a \emph{displaced thermal state}. Its ergotropy is $\mathcal W_b=\omega_b|\bm e_b^\dagger\bm\alpha|^2$ and its passive energy is $E_b^{\rm pas}=\omega_b\bm e_b^\dagger C\bm e_b$, where $\bm e_b$ is the unit vector that selects the battery mode.

We now specialize these general results to our charger--battery model. Importantly, both the coherent drive and the incoherent gain act through the same charger mode $a$
\[
\bm f=-i\mathcal E\bm e_a, \qquad K=\kappa_{a,\uparrow}\bm e_a\bm e_a^\dagger,
\]
where $\bm e_a$ selects the charger mode. To characterize how the excitations propagate from the charger to the battery, we introduce the \emph{impulse response} 
\begin{equation}
h_{ba}(u)=\bm e_b^\dagger e^{Mu}\bm e_a.
\label{eq:hba_SM}
\end{equation}
Substituting the above source and noise matrices into Eq.~\eqref{eq:sm_source_noise_integrals}, we obtain
\begin{equation}
\begin{aligned}
\frac{\mathcal W_b(t)}{\omega_b}
&=|\mathcal E|^2\left|\int_0^t h_{ba}(u)\,du\right|^2,\\
\frac{E_b^{\rm pas}(t)}{\omega_b}
&=\kappa_{a,\uparrow}\int_0^t|h_{ba}(u)|^2\,du.
\end{aligned}
\label{eq:sm_source_noise_exact}
\end{equation}
These two expressions reveal a fundamental difference between coherent driving and incoherent gain: the coherent-drive contributions are summed as complex amplitudes and can therefore interfere constructively or destructively, whereas the fluctuations injected by the gain channel accumulate independently over time.

These expressions are \emph{exact at any finite time} for vacuum initial states, even if the drift is unstable, has degenerate growth rates, or cannot be diagonalized. The collective reservoir modifies the response \(h_{ba}\) through the drift matrix \(M\), but, because it is a vacuum reservoir coupled through annihilation operators, it does not contribute directly to the normally ordered noise-injection matrix \(K\).

\suppsubsection{From temporal response to stored-work quality}

To quantify how effectively the coherent drive generates extractable work relative to the gain-induced fluctuations, we consider the work-quality ratio $R_b(t)=\mathcal W_b(t)/E_b^{\rm pas}(t)$, defined for $t>0$ with nonzero passive battery energy. For vacuum initial conditions, combining the two expressions in Eq.~\eqref{eq:sm_source_noise_exact} yields the exact \emph{finite-time source--noise relation}
\begin{equation}
\begin{aligned}
R_b(t)&=\frac{|\mathcal E|^2t}{\kappa_{a,\uparrow}}\mathcal M_b(t),\\
\mathcal M_b(t)&=
\frac{\big|\int_0^t h_{ba}(u)\,du\big|^2}
{t\int_0^t|h_{ba}(u)|^2\,du},\qquad
0\leq\mathcal M_b(t)\leq1.
\end{aligned}
\label{eq:sm_source_noise_bound}
\end{equation}
%
The physical meaning of $\mathcal{M}_b$ becomes clearer by introducing the time-averaged complex response and its mean squared magnitude,
\begin{equation}
\overline h_{ba}(t)=\frac{1}{t}\int_0^t h_{ba}(u)\,du,
\quad
\overline{|h_{ba}|^2}(t)=\frac{1}{t}\int_0^t |h_{ba}(u)|^2\,du.
\label{eq:sm_response_time_averages}
\end{equation}
The temporal matching factor can then be written as
\begin{equation}
\mathcal M_b(t)=
\frac{|\overline h_{ba}(t)|^2}{\overline{|h_{ba}|^2}(t)}
=1-\frac{\overline{|h_{ba}-\overline h_{ba}|^2}(t)}{\overline{|h_{ba}|^2}(t)}.
\label{eq:sm_matching_variance_form}
\end{equation}
This expression shows that \emph{temporal variations in the complex response reduce \(\mathcal M_b\)}. The Cauchy--Schwarz inequality gives $0\leq\mathcal M_b\leq1$ whenever the denominator is nonzero, with equality at unity only if the complex response is constant almost everywhere on $[0,t]$. Therefore, strong temporal matching requires both the amplitude and phase of the response to remain nearly constant. The resulting bound $R_b(t)\leq |\mathcal E|^2t/\kappa_{a,\uparrow}$ sets an upper limit on the ergotropy relative to the passive energy at a given time. The prefactor in Eq.~\eqref{eq:sm_source_noise_bound} also shows how the gain rate and charging time affect the work-quality ratio.


We now relate $R_b$ to the work-like SNR at a prescribed target occupation $n_\star$. At the first-passage time $t_{\rm ch}(n_\star)$, define $w=\mathcal W_b(t_{\rm ch})/\omega_b$ and $q=n_b^{\rm inc}(t_{\rm ch})=E_b^{\rm pas}(t_{\rm ch})/\omega_b$. The total battery occupation is $n_\star=w+q$ and the work-quality ratio gives $R_b=w/q$. Thus, at this stopping time,
\begin{equation}
w=\frac{n_\star R_b}{1+R_b},\qquad
q=\frac{n_\star}{1+R_b}.
\label{eq:sm_matching_wq}
\end{equation}
Substituting Eq.~\eqref{eq:sm_matching_wq} into Eq.~\eqref{eq:sm_energy_variance_displaced} gives
\begin{equation}
\mathrm{SNR}_{b,\star}=
\frac{\sqrt{n_\star}\,R_b}
{\sqrt{(1+R_b)^2+n_\star(1+2R_b)}}.
\label{eq:sm_matching_quality}
\end{equation}
Equation~\eqref{eq:sm_matching_quality} shows that the work-like SNR increases monotonically with \(R_b\) at fixed \(n_\star\). This can be verified explicitly by expressing the SNR in terms of the passive occupation \(q\):
\begin{equation}
\mathrm{SNR}_{b,\star}=
\frac{n_\star-q}{\sqrt{n_\star+2n_\star q-q^2}},
\qquad
\frac{d\mathrm{SNR}_{b,\star}}{dq}
=-\frac{n_\star(n_\star+1)}{(n_\star+2n_\star q-q^2)^{3/2}}<0.
\label{eq:sm_matching_monotonic}
\end{equation}
Thus, reducing the passive occupation at a fixed target increases the work-like SNR. Together with the finite-time source–noise relation, this establishes a direct connection between the temporal response of the coupled modes and the quality of the stored energy.

\suppsubsection{Mode interference and drive-frequency dependence}

We now use the finite-time response derived above to examine how mode interference affects the stored-work quality and how driving-frequency optimization modifies the coherent response. To connect the response with the relevant dynamical modes, suppose that the drift matrix $M$ is diagonalizable, with right and left eigenvectors $\bm v_j$ and $\bm u_j$, normalized such that $\bm u_j^\dagger\bm v_k=\delta_{jk}$. The response can then be expressed as a sum over all dynamical modes:
\begin{equation}
h_{ba}(u)=\sum_j A_j e^{\lambda_j u},\qquad
A_j=(\bm e_b^\dagger\bm v_j)(\bm u_j^\dagger\bm e_a).
\label{eq:sm_matching_modal_response}
\end{equation}
Substituting this expansion into the expressions of the ergotropy and passive energy shows how the dynamical modes contribute differently to the coherent-drive and gain-noise responses. The coherent displacement depends on \[
\big|\sum_j A_j\int_0^t e^{\lambda_j u}\,du\big|^2,
\]
while the noise contribution depends on \[
\sum_{j,k}A_j A_k^*\int_0^t e^{(\lambda_j+\lambda_k^*)u}\,du.
\]
Both expressions contain interference terms between different modes. The coherent response sums amplitudes over time before taking the modulus squared, while the noise contribution integrates the squared response magnitude. This distinction underlies the temporal response matching discussed above.

A simpler expression emerges when a single unstable mode dominates the long-time response. If an eigenmode with $\lambda_+=\gamma+i\Omega$ has a substantially larger real part than all other modes, with $\gamma>0$ and nonzero source--battery overlap, the work-quality ratio approaches
\begin{equation}
\lim_{t\to\infty} R_b(t)
=\frac{2\gamma}{|\lambda_+|^2}
\frac{|\bm u_+^\dagger\bm f|^2}
{\bm u_+^\dagger K\bm u_+}
=\frac{2\gamma|\mathcal E|^2}
{\kappa_{a,\uparrow}|\lambda_+|^2}.
\label{eq:sm_matching_single_mode}
\end{equation}
The second equality follows because the coherent drive and gain noise enter through the same charger port. Their projections onto the dominant mode share a common factor that cancels in the ratio. Therefore, the asymptotic work quality depends on the dominant eigenvalue and the drive-to-gain ratio. When two modes share the largest real part, their interference can persist at long times and both contributions must be retained.

We next study how the driving frequency modifies this temporal accumulation. The bare frequencies of the charger and battery are fixed at $\omega_a$ and $\omega_b$. We define $\delta_d=(\omega_a+\omega_b)/2-\omega_d$ and $\Delta=(\omega_a-\omega_b)/2$. In all charging benchmarks, we set $\omega_a=\omega_b$ ($\Delta=0$). Resonant driving corresponds to $\delta_d=0$, while frequency optimization of the coherent scheme varies only $\delta_d$.

%

In the rotating frame at the driving frequency, the diagonal detunings are $\delta_d+\Delta$ and $\delta_d-\Delta$. Keeping all other system parameters fixed, changing the driving frequency therefore shifts the drift matrix by a purely imaginary multiple of the identity, i.e., $M(\delta_d)=M(0)-i\delta_d\mathbb I$. Consequently, the eigenvalues acquire the same imaginary shift, while their real parts and hence the instability growth rates remain unchanged. The impulse response and incoherent battery occupation satisfy
\begin{equation}
h_{ba}(u;\delta_d)=e^{-i\delta_d u}h_{ba}(u;0),\qquad\qquad
n_b^{\rm inc}(t;\delta_d)=n_b^{\rm inc}(t;0).
\label{eq:sm_matching_drive_offset}
\end{equation}
Thus, tuning the driving frequency changes how the coherent response accumulates over time but leaves the gain-induced passive occupation unchanged at any fixed time. 

This distinction is particularly relevant for coherent coupling. At the matched operating point used in the main text, the two drift eigenvalues, in units of \(\kappa_0\), are $\lambda_\pm/\kappa_0=0.05\pm 3.15i$. The corresponding resonant-drive impulse response is $h_{ba}^{\rm coh}(u)=-i(J^*/\Omega)e^{0.05u}\sin(\Omega u)$, with $\Omega=3.15$, where time and frequency are expressed in units of \(\kappa_0^{-1}\) and \(\kappa_0\), respectively. Its alternating sign causes successive portions of the resonantly driven response to interfere \emph{destructively}, which strongly suppresses the accumulated coherent displacement. Tuning the driving frequency toward either hybrid-mode frequency reduces this cancellation by preferentially enhancing one frequency component. Nevertheless, both oscillatory modes remain present in the response.

The dissipative-coupling scheme exhibits a qualitatively different response. At its matched operating point, the eigenvalues are $\lambda_+/\kappa_0=0.05$ and $\lambda_-/\kappa_0=-8.98$. The corresponding impulse response is proportional to $h_{ba}^{\rm diss}\propto e^{0.05u}-e^{-8.98u}$, which has a constant complex phase and a positive real envelope for $u>0$. Unlike the coherent case, contributions from different times add \emph{constructively} under resonant driving.

This fixed response phase makes resonant driving optimal for the dissipative-coupling scheme. By the triangle inequality,
\[
\left|
\int_0^t e^{-i\delta_d u}h_{ba}(u;0)\,du
\right|
\leq
\int_0^t|h_{ba}(u;0)|\,du
=
\left|
\int_0^t h_{ba}(u;0)\,du
\right|.
\]
Thus, the coherent battery occupation is maximized at \(\delta_d=0\) for every fixed time. Since the passive occupation is independent of \(\delta_d\), resonant driving also maximizes the total battery occupation at each time and therefore minimizes the first-passage time for any reachable target in this model.


At $n_\star=1000$, the temporal matching factor $\mathcal M_b$ is approximately $4.9\times10^{-4}$ for the resonantly-driven coherent scheme, $0.5$ for frequency-optimized coherent scheme, and $0.98$ for the dissipative scheme. The corresponding work-like SNRs are approximately $14.8$ and $7.6$ for the dissipative and optimized coherent schemes, respectively. The dissipative scheme benefits from stronger temporal matching and a lower gain rate.


\suppsubsection{First-passage optimization of coherent charging}

We now derive the optimal driving frequency for the coherent-coupling scheme at a prescribed target occupation $n_\star$. We set $\Gamma_{\rm col}=0$ and $\Delta=0$, while retaining a finite incoherent gain $\kappa_{a,\uparrow}$. Starting from the master equation in the main text, the first moments $\alpha=\langle a\rangle$ and $\beta=\langle b\rangle$ obey
\begin{equation}
\begin{aligned}
\dot\alpha&=-(\gamma_a+i\delta_d)\alpha-iJ\beta-i\mathcal E,\\
\dot\beta&=-(\gamma_b+i\delta_d)\beta-iJ\alpha,
\end{aligned}
\label{eq:sm_coh_opt_first_moments}
\end{equation}
where $\gamma_a=(\kappa_{a,\downarrow}-\kappa_{a,\uparrow})/2$ and $\gamma_b=\kappa_b/2$. We choose $J$ real, without loss of generality. Defining $\lambda=-(\gamma_a+\gamma_b)/2$, $\mu=(\gamma_a-\gamma_b)/2$, and $\Omega=\sqrt{J^2-\mu^2}$, we consider the strong-hybridization regime $J^2>\mu^2$ relevant to our benchmark. The charger-to-battery impulse response follows by exponentiating the drift matrix of Eq.~\eqref{eq:sm_coh_opt_first_moments}:
\begin{equation}
h_{ba}(u;\delta_d)=-i\frac{J}{\Omega}
 e^{(\lambda-i\delta_d)u}\sin(\Omega u).
\label{eq:sm_coh_opt_kernel}
\end{equation}
For vacuum initial conditions, the coherent displacement and the normally ordered fluctuations follow from the source--noise solutions in Eq.~\eqref{eq:sm_source_noise_integrals}. We introduce the complex response integral
\begin{equation}
\begin{aligned}
I(t,\delta_d)&=\int_0^t e^{(\lambda-i\delta_d)u}\sin(\Omega u)\,du\\
&=\frac{1}{2i}\left(
\frac{e^{z_+t}-1}{z_+}-\frac{e^{z_-t}-1}{z_-}\right),\\
z_\pm&=\lambda-i(\delta_d\mp\Omega).
\end{aligned}
\label{eq:sm_coh_opt_response_integral}
\end{equation}
The battery occupation then separates exactly into coherent and incoherent parts:
\begin{equation}
\begin{aligned}
\langle n_b(t;\delta_d)\rangle
&=\frac{|\mathcal E|^2J^2}{\Omega^2}|I(t,\delta_d)|^2+q(t),\\
q(t)&=\frac{\kappa_{a,\uparrow}J^2}{\Omega^2}
\int_0^t e^{2\lambda u}\sin^2(\Omega u)\,du.
\end{aligned}
\label{eq:sm_coh_opt_occupation}
\end{equation}
Here $q(t)=n_b^{\rm inc}(t)$ is the incoherent battery occupation arising from gain-induced fluctuations. In particular, $q(t)$ does not depend on $\delta_d$, and $q'(t)\geq0$. Thus, only the coherent displacement changes when the driving frequency is varied at fixed time.

For a prescribed target $n_\star$, we define the \emph{optimized} first-passage time as
\begin{equation}
t_\star=\min_{\delta_d}t_{\rm ch}(n_\star;\delta_d).
\end{equation}
At a smooth optimum $(t_\star,\delta_\star)$ with a transversal target crossing, implicit differentiation gives $dt_{\rm ch}/d\delta_d=-(\partial_{\delta_d}\langle n_b\rangle)/(\partial_t\langle n_b\rangle)$. Hence the candidate optimum satisfies
\begin{equation}
\begin{aligned}
n_\star&=q(t_\star)+\frac{|\mathcal E|^2J^2}{\Omega^2}
|I(t_\star,\delta_\star)|^2,\\
0&=\operatorname{Im}\!\left[I^*(t_\star,\delta_\star)
I_1(t_\star,\delta_\star)\right],\\
I_1(t,\delta_d)&=\int_0^t u\,e^{(\lambda-i\delta_d)u}
\sin(\Omega u)\,du.
\end{aligned}
\label{eq:sm_coh_opt_first_passage}
\end{equation}
The second condition follows from $\partial_{\delta_d}I=-iI_1$. The optimal driving frequency is determined by comparing the first-passage times across all relevant charging branches. 

For $0<t\leq\pi/\Omega$, the kernel $e^{\lambda u}\sin(\Omega u)$ has a fixed sign, so the triangle inequality makes $\delta_d=0$ optimal at each fixed time. When the target exceeds the maximum occupation attainable during this first oscillation interval, charging must continue into a later interval. An off-resonant drive can then become optimal.


At a fixed target, the ergotropy at the first-passage time is $\mathcal W_b=\omega_b[n_\star-q(t_{\rm ch})]$. Since $q(t)$ is nondecreasing, minimizing the first-passage time also maximizes the ergotropy, average ergotropy power, and work-like SNR among the considered driving frequencies.

The connection to the stationary optimization of Ref.~\cite{Ahmadi2025} can be made explicit. If $\lambda<0$, the response has the steady-state limit $I(\infty,\delta_d)=\Omega/[(\lambda-i\delta_d)^2+\Omega^2]$. Maximizing the stationary coherent occupation then gives
\begin{equation}
\delta_{d,{\rm opt}}^2=\Omega^2-\lambda^2
=J^2-\frac{(\kappa_{a,\downarrow}-\kappa_{a,\uparrow})^2+\kappa_b^2}{8},
\label{eq:sm_coh_opt_stationary_limit}
\end{equation}
provided the right-hand side is positive; otherwise the optimum is $\delta_d=0$. Setting $\kappa_{a,\uparrow}=0$ recovers the purely lossy result in Ref.~\cite{Ahmadi2025}. Our unstable benchmark has \(\lambda/\kappa_0=0.05>0\), so its optimal driving frequency is determined by the finite-time first-passage analysis above.

\suppsubsection{Robustness of the work-like SNR advantage}

We next examine the robustness of this SNR advantage against variations in the coupling strengths. We vary
$J$ and $\Gamma_{\rm col}$ independently, while keeping
the gain rates fixed at their respective reference values.
For each $J$, we optimize the
driving frequency to minimize the first-passage time
at $n_\star=1000$.

\begin{figure}[t]
\centering
\includegraphics[width=0.45\columnwidth]{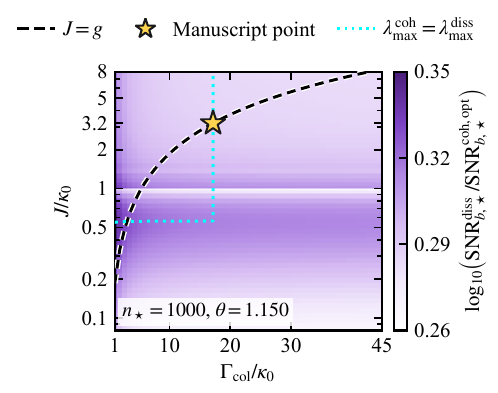}
\caption{Robustness of the dissipative work-like SNR advantage against coupling-strength variations.
The color scale shows
\(\log_{10}(\mathrm{SNR}_{b,\star}^{\rm diss}/\mathrm{SNR}_{b,\star}^{\rm coh,opt})\)
at $n_\star=1000$, with $J$ and $\Gamma_{\rm col}$
varied independently and $\theta=1.15$ fixed.
The driving frequency is optimized in the coherent-coupling scheme (for each $J$) to minimize
the first-passage time, while the dissipative-coupling scheme uses
resonant driving. The black dashed curve marks the matched-coupling line
$J=g$, the cyan dotted curve marks the equal maximal growth rates, and the star marks the reference operating
point used in the main text. The gain rates are fixed at
$\kappa_{a,\uparrow}/\kappa_0=2.2$ (coherent)
and $1.3046$ (dissipative), so the growth rates
are not generally matched across the scan. The remaining parameters are
$\kappa_{a,\downarrow}/\kappa_0
=\kappa_b/\kappa_0=1$,
$\mathcal E/\kappa_0=10$, and
$\Delta=0$. The sharp horizontal feature near \(J/\kappa_0\simeq1\) originates from a transition between optimal first-passage times in the coherent scheme.
\justifying}
\label{fig:snr_landscape}
\end{figure}

Figure~\ref{fig:snr_landscape} shows that dissipative
coupling maintains a higher work-like SNR than the frequency-optimized coherent scheme, i.e., 
\[
\mathrm{log}_{10}\left(\mathrm{SNR}_{b,\star}^{\rm diss}/\mathrm{SNR}_{b,\star}^{\rm coh,opt} \right)>0
\]
throughout the investigated parameter range. 
This advantage extends well beyond the matched-coupling line \(J=g\), demonstrating its robustness across the scanned coupling strengths.
The sharp horizontal feature near \(J/\kappa_0\simeq1\) results from a step-like change in the optimal first-passage time $t_{\rm ch}(n_\star)$. At smaller \(J\), resonant driving is optimal. As $J$ increases, the first battery-occupation
peak decreases and eventually falls below the target
$n_\star=1000$. The optimized coherent scheme then reaches this target at a later time with an off-resonant drive.
The longer charging time allows more gain-induced fluctuations to accumulate, reducing the work-like SNR at the same target occupation. 


%
\begin{figure*}[t]
\centering
\includegraphics[width=0.7\textwidth]{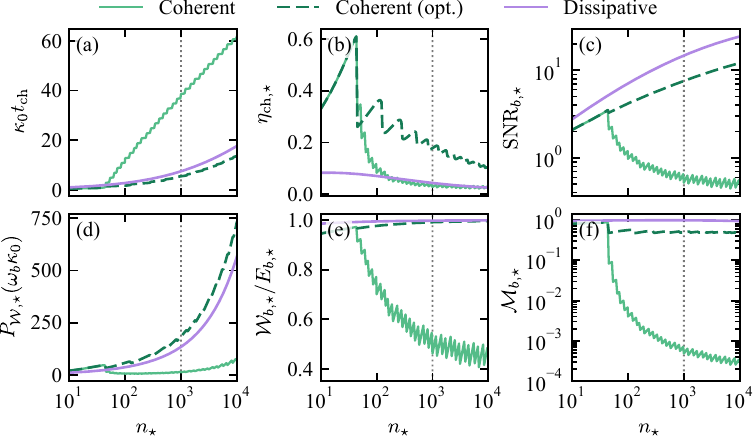}
\caption{Extended target-occupation charging performance.
First-passage results for $10\leq n_\star\leq 10^4$.
(a) Charging time $\kappa_0 t_{\mathrm{ch}}$.
(b) Signed-input charging efficiency $\eta_{\mathrm{ch},\star}$.
(c) Work-like SNR $\mathrm{SNR}_{b,\star}=\mathcal W_{b,\star}/\sigma_{E,b,\star}$.
(d) Average ergotropy power $P_{\mathcal{W},\star}/(\omega_b\kappa_0)$.
(e) Ergotropy fraction $\mathcal{W}_{b,\star}/E_{b,\star}$.
(f) Temporal matching factor $\mathcal{M}_{b,\star} = \mathcal{M}_b[t_{\mathrm{ch}}(n_\star)]$.
The green solid, dark-green dashed, and purple solid curves represent the resonantly-driven coherent, frequency-optimized coherent, and dissipative schemes, respectively.  The vertical dotted lines mark the reference target $n_\star=1000$.
All parameters are the same as those in Fig.~4 of the main text. \justifying}
\label{fig:SM_extended_capacity}
\end{figure*}

\suppsubsection{Extension to larger target occupations}

Finally, we extend the finite-capacity comparison in Fig.~4 of the main text to larger target occupations (up to $n_\star=10^4$). We retain the same linear Gaussian model, operating parameters, and matched maximal growth rates. As in the main text, for each target, the charging time is defined as the earliest time at which the total battery occupation reaches $n_\star$. For the coherent-coupling scheme, we also optimize the driving frequency to minimize this first-passage time.

As shown in Fig.~\ref{fig:SM_extended_capacity}, the main trends persist over the extended target range. The resonantly driven coherent scheme continues to exhibit
beating-induced charging delays, while the frequency optimization largely removes them. The optimized coherent scheme charges faster and achieves higher average ergotropy power, while the dissipative-coupling scheme maintains a substantially higher work-like SNR, primarily due to its larger temporal matching factor $\mathcal{M}_{b,\star}$. Moreover, the dissipative-coupling scheme achieves a larger ergotropy fraction over most of the investigated target range. These results are obtained within the unsaturated linear-gain model. In the finite-emitter implementation discussed in Note~\ref{sec:saturable_gain}, gain saturation can limit the attainable stationary battery occupation.

\end{document}